\pdfoutput=1
\documentclass[useAMS,usenatbib,fleqn]{mn2e}
\usepackage{journals}
\usepackage{graphicx,color}
\usepackage{epstopdf}
\usepackage[latin1]{inputenc}
\usepackage[T1]{fontenc}
\usepackage{ae,aecompl}
\usepackage{graphics}
\usepackage{amsfonts}
\usepackage{amsmath}
\usepackage{multicol}
\usepackage{layout}
\usepackage{amssymb}
\usepackage[english]{babel}
\usepackage[a4paper,colorlinks=true,pdfstartview=FitV,linkcolor=red,citecolor=blue,urlcolor=magenta,draft]{hyperref}
 
\usepackage{times}

\newcommand\reallywidehat[1]{\arraycolsep=0pt\relax%
\begin{array}{c}
\stretchto{
  \scaleto{
    \scalerel*[\widthof{\ensuremath{#1}}]{\kern-.5pt\bigwedge\kern-.5pt}
    {\rule[-\textheight/2]{1ex}{\textheight}} 
  }{\textheight} %
}{0.5ex}\\           
#1\\                 
\rule{-1ex}{0ex}
\end{array}
}

\title[Shapes  and overdensities]  {A look
  to the inside  of haloes: a characterisation of the halo shape  as a function
  of   overdensity in the Planck cosmology}
\author[Despali    et     al.     2016]{\parbox{\textwidth}{
    Giulia Despali$^{1}$\thanks{E-mail:
      \href{mailto:gdespali@gmail.com}   {gdespali@gmail.com}},
    Carlo Giocoli$^{2,3}$,
    Mario  Bonamigo$^{2,4}$,
    Marceau  Limousin$^{2}$,
    Giuseppe  Tormen$^3$\\}\\
  $^1$  Max  Planck   Institute  for
  Astrophysics, Karl-Schwarzschild-Strasse 1,  85740 Garching, Germany
  \\
  $^2$  Aix  Marseille   Universit\'e,  CNRS,  LAM  (Laboratoire
  d'Astrophysique  de Marseille)  UMR 7326,  13388, Marseille,  France
  \\
  $^3$Dipartimento di Fisica e Astronomia, Alma Mater Studiorum Universit\`{a} di 
Bologna, viale Berti Pichat, 6/2, 40127 Bologna, Italy\\
$^4$  Dark Cosmology Centre, Niels Bohr Institute,
 University of Copenhagen, Juliane Maries Vej 30, DK-2100 Copenhagen, Denmark\\
  $^5$  Dipartimento di Fisica e Astronomia,
  Universit\`{a} degli  Studi di  Padova, vicolo  dell'Osservatorio 3,
  35122, Padova, Italy
}

\begin{document}
\date{}
\maketitle
\label{firstpage}
\pagerange{\pageref{firstpage}--\pageref{lastpage}} \pubyear{2016}
\begin{abstract}

  In this paper we study the triaxial properties of dark matter haloes
  of a wide range of masses extracted from a set of cosmological
  $N$-body simulations.  We measure the shape at different distances
  from the halo centre (characterised by different overdensity
  thresholds), both in three and in two dimensions. We
    discuss how halo triaxiality increases with mass, redshift and
  distance from the halo centre. We also examine how the
  orientation of the different ellipsoids are aligned with each other
  and what is the gradient in internal shapes for halos with different
  virial configurations.  Our findings highlight that the internal
  part of the halo retains memory of the violent formation process
  keeping the major axis oriented toward the preferential direction of
  the in-falling material while the outer part becomes rounder due to
  continuous isotropic merging events.  This effect is clearly evident
  in high mass haloes - which formed more recently - while it is more
  blurred in low mass haloes.  We present simple distributions that
  may be used as priors for various mass reconstruction algorithms,
  operating in different wavelengths, in order to recover a more
  complex and realistic dark matter distribution of isolated and
  relaxed systems.

\end{abstract}

\section{Introduction}

Different wide field surveys, observing at various wavelengths, are
revealing that most of the matter density content of our Universe is
in form of collisionless particles \citep{amara12,guzzo14,covone14}.
These particles do not emit radiation and interact only
gravitationally with the surrounding density field: they are generally
termed Dark Matter.
Following the standard scenario of structure formation, dark matter
drives the structure evolution processes:  systems up
to proto-galactic scales form as consequence of gravitational collapse
and then merge together, along the cosmic time, forming the more
massive ones \citep{white78,tormen98a,springel01b,tormen04}.  Galaxy
clusters sit at the top of this hierarchical pyramid being the most
massive and late forming virialized structures of our Universe
\citep{frenk90,borgani11,angulo12}.

Various studies of time evolving isolated perturbations seeding in the
dark matter density field have lead many scientist to the development
of the spherical collapse model
\citep{white79,press74,pace10}.  A density perturbation grows with the
expansion of the Universe in concentric shells and, if it is dense
enough, will pull away from the background expansion, and will
collapse after reaching a maximum size characterised by null kinetic
energy.  The collapse happens when the perturbation exceeds the
predicted critical value by the spherical collapse model forming a so
called dark matter halo \citep{bond91,eke96,bryan98}.
It is  interesting to notice that  density perturbations from
which haloes  form are not  independent with each other  nor isolated,
but   during   the   expansion    and   collapse   perturbations   are
typically pulled,  stretched and  sheared by  the surrounding
density  field   \citep{doroshkevich70,despali13}.   All   these  also
translate   in   a  mass   dependence   of   the  collapse   threshold
\citep{eisenstein95,sheth99b,sheth01b,sheth02} and in the formation of
haloes  that  are  generally  triaxial,  more  in  particular  prolate
\citep{jing02,despali14,bonamigo15,vega16}.    Thus,    the   standard
spherical modelling of the dark  matter, stars, inter-galactic and the
intra-cluster medium is  only a rough approximation  and in particular
both theory and observations agree  on the general picture that haloes
in which  galaxies and clusters live  are very well approximated  by a
triaxial                                                     ellipsoid
\citep{morandi11a,morandi12,sereno12,limousin13,groener14}.  The study
of the  asphericity of  galaxy clusters  is growing,  in light  of the
analyses performed on different  numerical simulations during the last
years \citep{angrick10,rossi10,rossi11,despali14,vega16}.

In this paper we aim to discuss the dependence of the halo shape on
the distance from the centre.  Other previous works
\citep{allgood06,bailin05} have measured shapes at different fractions
of the virial radius.  For reasons that are linked to the different
halo mass definitions, $M_{500}$, $M_{200}$ or $M_{vir}$
\citep{despali16} and that we will better discuss
later in the text,
in this work we choose to present our
results in term of different overdensity thresholds: we define halos
as triaxial regions enclosing a desired multiple of the critical
density of the Universe and, for each halo identified using
$\Delta_{vir}$, we measure the triaxial shape at other four
overdensity thresholds, multiples of the critical density of the
Universe $\rho_c$ (200, 500, 1000 and 2000).  
The choice of these values is also
motivated by the fact that, typically, the X-ray community adopts
a mass definition that is associated
to the region enclosing 500 times the
critical density, while weak lensing and dynamical analyses usually
prefer 200 times the critical density.  On the other side, strong
lensing researchers make use of the very central region of clusters or
galaxies, where critical lines emerge; in this case we can refer to
the  region enclosing at least 1000 times the critical density of
the Universe \citep{broadhurst05b,coe12,zitrin11a}.  In light of the
various observational analyses, it is important to underline that a
study of the degree of alignment of the mass density distribution at
different distances from the centre can help us to understand dark
matter and baryonic properties in which the various physical processes
are taking place.  \citet{hopkins05} have studied the shape properties
of haloes extracted from a light-cone up to redshift $z=3$ constructed
from a large-scale high-resolution $N$-body simulation.  They discuss
that the mean halo ellipticity increases with redshift as $\langle
\epsilon \rangle = 0.33 + 0.05 z$ and with the cluster mass, as also
found by \citet{despali14,bonamigo15}.  On the other hand, for
\citet{hopkins05} the cluster ellipticity decreases with radius in
disagreement with other results
\citep{bailin05,hayashi07,vera-ciro11,velliscig15}.  In particular
\citet{vera-ciro11}, studying the $N$-body haloes from the Aquarius
simulation \citep{springel08b}, found that the evolution in halo shape
correlates well with the distribution of the in-falling material:
prolate configurations arise when haloes are fed through narrow
filaments whereas triaxial/oblate configurations result as the
accretion turns more isotropic at later times.  The geometrical
properties of haloes at different epochs are not lost: haloes retain
memory of their structure at earlier times. This memory is imprinted
in their present-day shape trends with radius, which change from
typically prolate in the inner (earlier collapsed) regions to a
triaxial in the outskirts -- corresponding to the shells that have
collapsed last and are now at the virial radius.

In this work we will present a study of the shape properties of haloes
extracted from DM only simulations.  We underline that our results do
not account for the presence of baryons -- mainly influencing the most
internal shells -- and could eventually be adapted to their presence
using pre-calibrated analytical recipes.

The paper is organised as follows: $(i)$ Section 2 describes the
numerical simulations and the halo catalogues; $(ii)$ in Section 3 we
discuss how we selected relaxed and regular halos; $(iii)$
our results are presented in Sections 4 and 5, which show respectively
the distributions derived in three or two dimensions; $(iv)$ Section 6
is dedicated to summarise our results and draw our conclusions.

\begin{figure}
\centering
\includegraphics[width=\hsize]{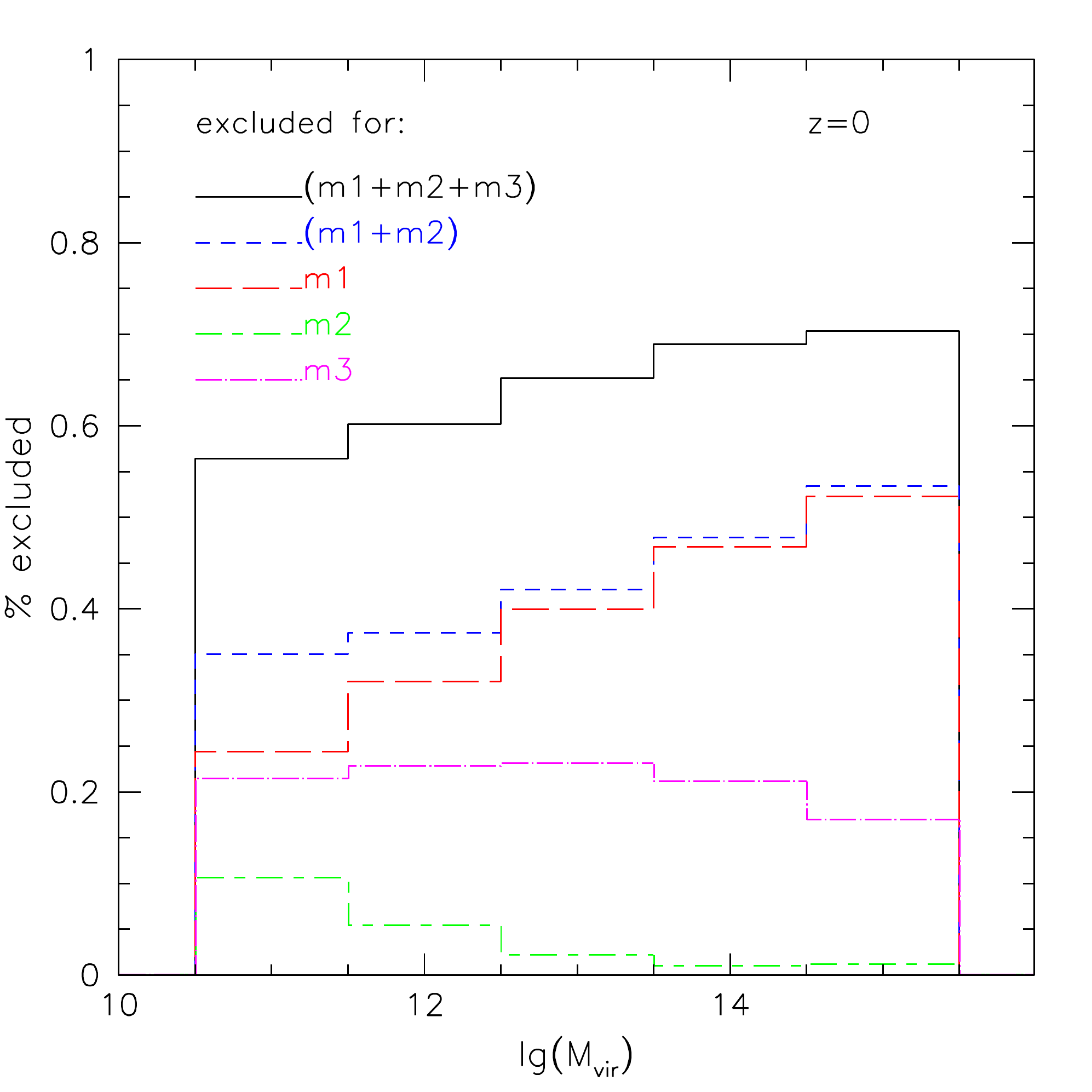}
\includegraphics[width=\hsize]{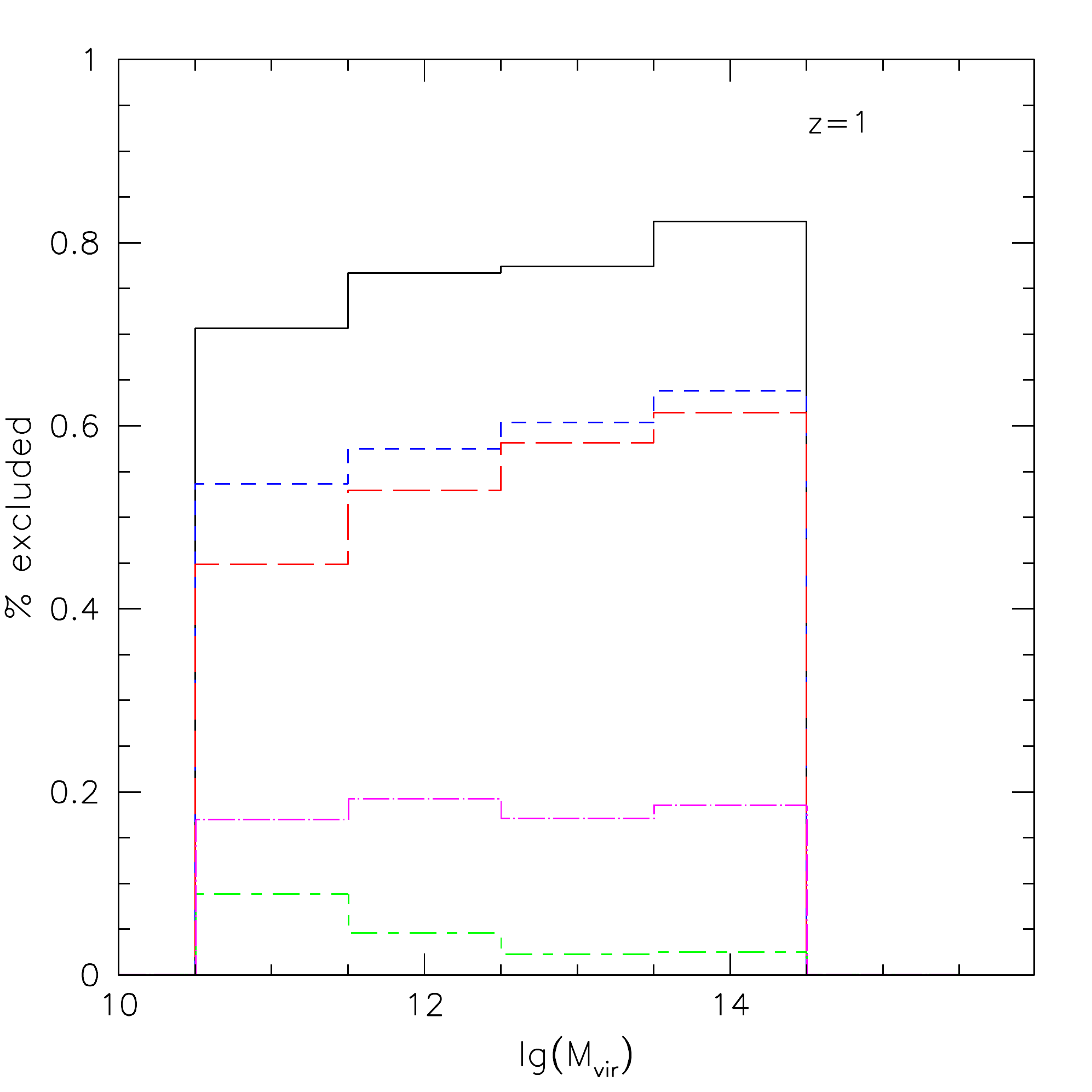}
\caption{Fraction of haloes excluded by each selection criterion - or
  by their combination - for different mass bins.  The red
    histogram shows the percentage of irregular or ``unrelaxed''
    haloes detected by \textit{method 1} (\textit{m1}), the green
    colour stands for those detected with \textit{method 2}
    (\textit{m2}), the association of the two criteria is shown in
    blue.  The fraction of haloes excluded with \textit{method 3}
    (\textit{m3}) is shown in magenta and finally the combination of
    all the three methods in black.  At $z=0$ (top panels), our
    remaining catalogue contains roughly 60 \% of haloes of
    $10^{12}M_{\odot}h^{-1}$ and 40\% of high mass haloes of
    $10^{15}M_{\odot}h^{-1}$; at $z=1$ (bottom panels) the percentage
    of selected haloes is further reduced.\label{sel_crit}}
\end{figure}

\section{The Numerical Simulations}\label{simulations}
\subsection{Le SBARBINE simulations}

Le SBARBINE simulations are six cosmological simulations which have
been run in Padova using the publicly available code GADGET-2
\citep{springel05a}; these are part of a series of new simulations
which have been presented in a previous work \citep{despali16}.  The
adopted cosmology follows the recent Planck results
\citep{planck1_14}, in particular we have set: $\Omega_{m}=0.307$,
$\Omega_{\Lambda}=0.693$, $\sigma_{8}=0.829$ and $h=0.677$.  The
initial power spectrum was generated with the code CAMB \citep{camb}
and initial conditions were produced perturbing a glass distribution
with N-GenIC (\url{http://www.mpa-garching.mpg.de/gadget}).  They all
follow $1024^{3}$ collisionless particles in a periodic box of
variable length (we refer the reader to Table~\ref{tab_sim} for more
details).

\begin{figure*} \centering
\includegraphics[width=0.33\hsize]{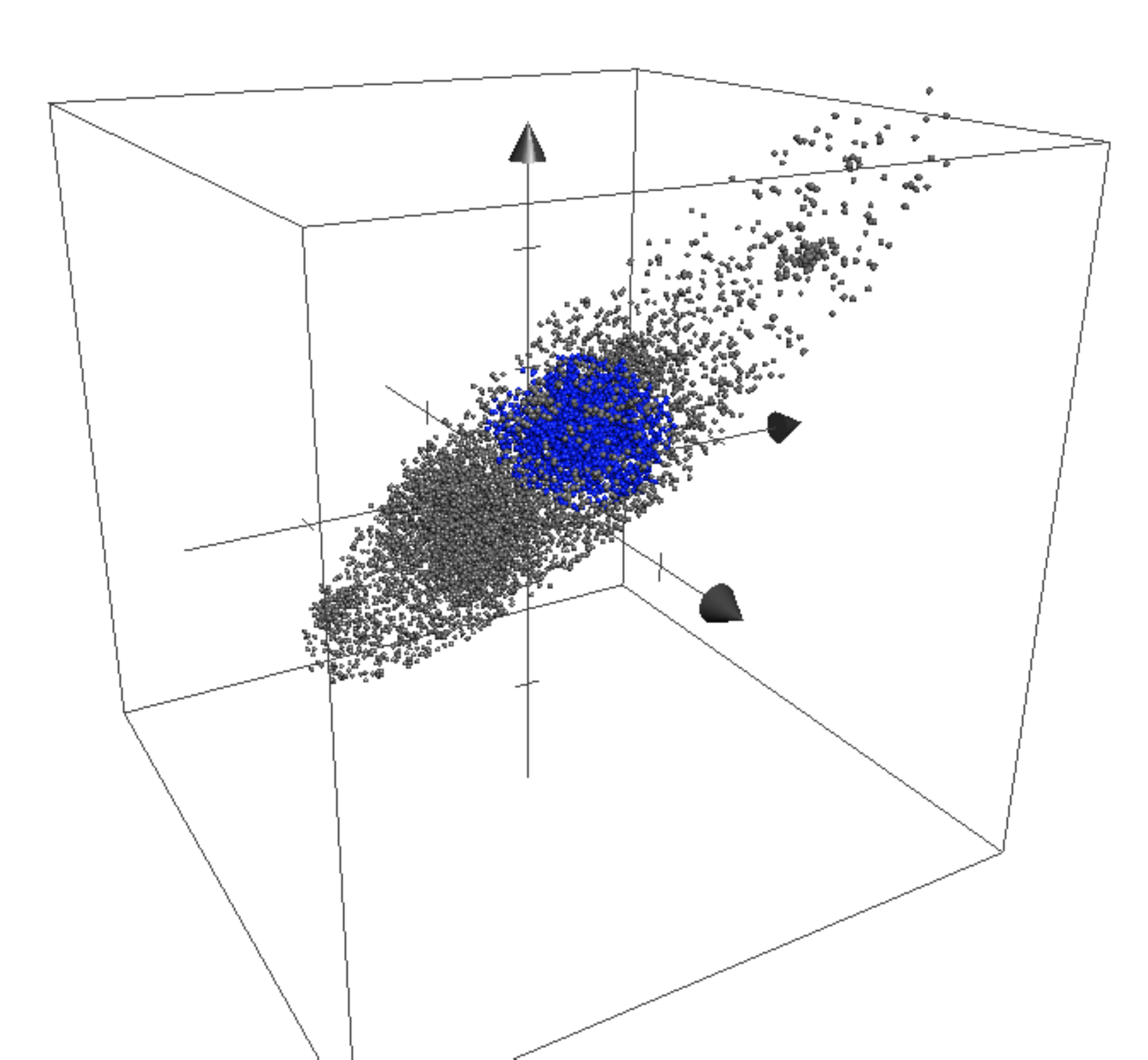}
\includegraphics[width=0.33\hsize]{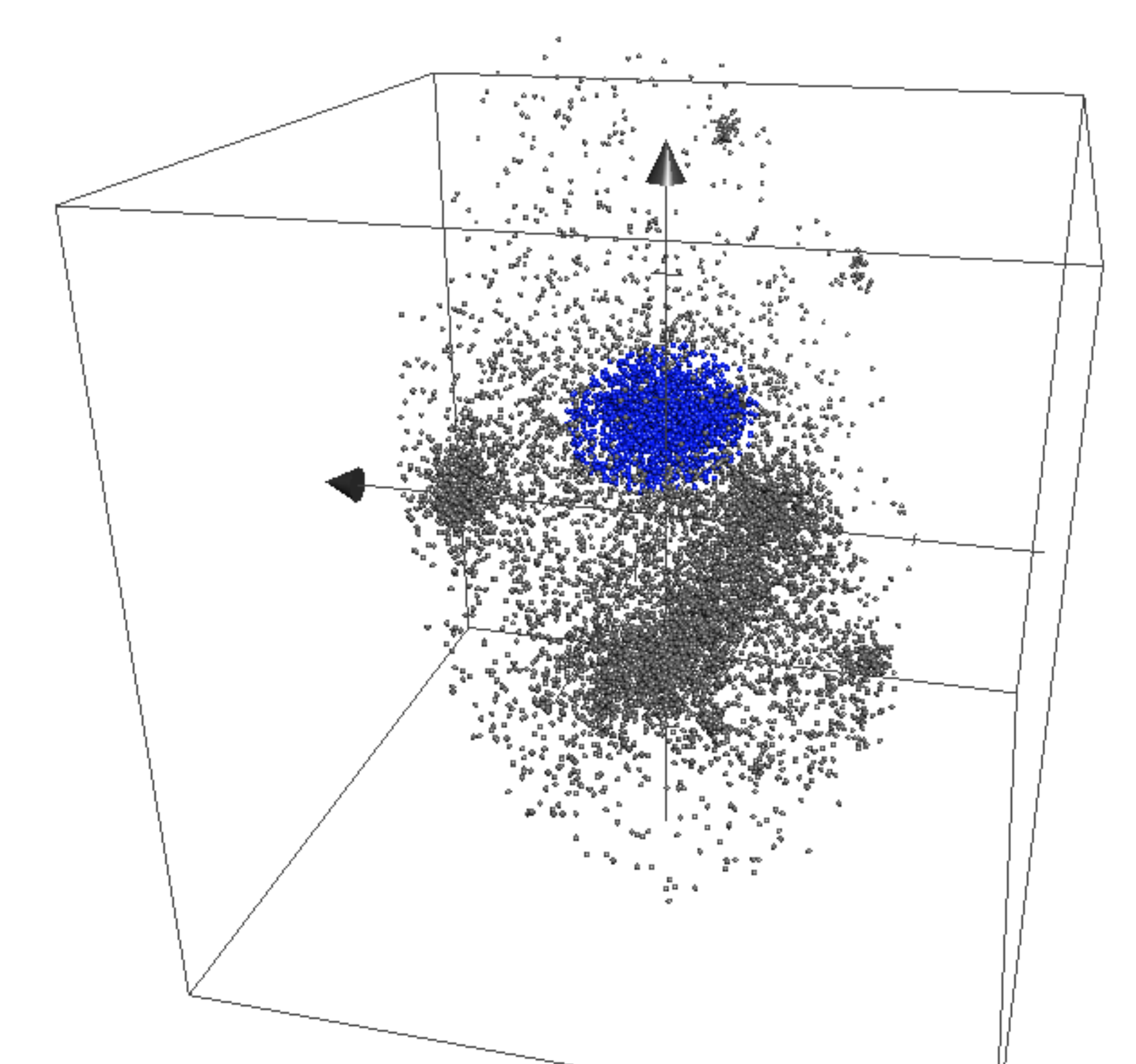}\\
\includegraphics[width=0.33\hsize]{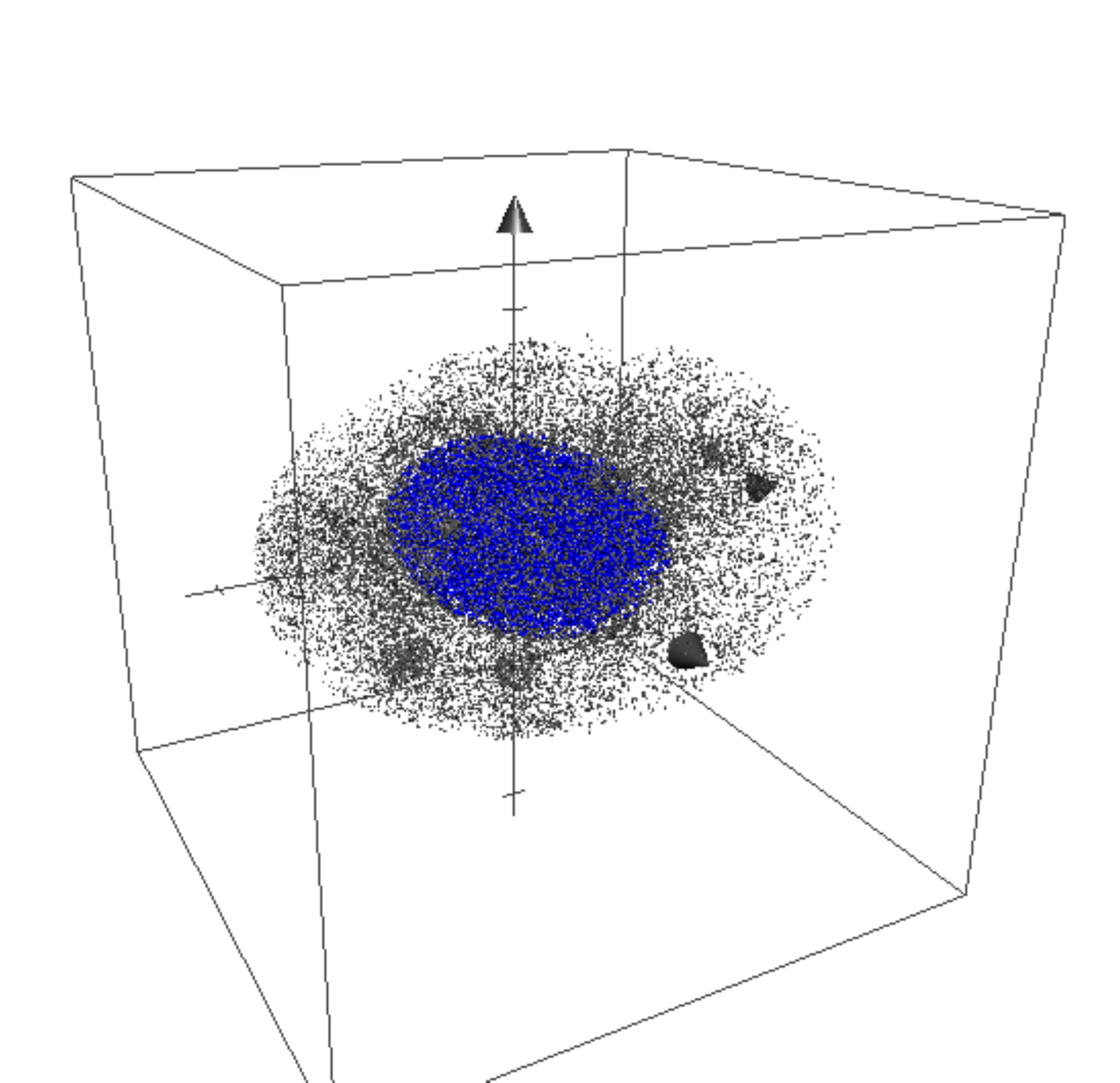}
\includegraphics[width=0.33\hsize]{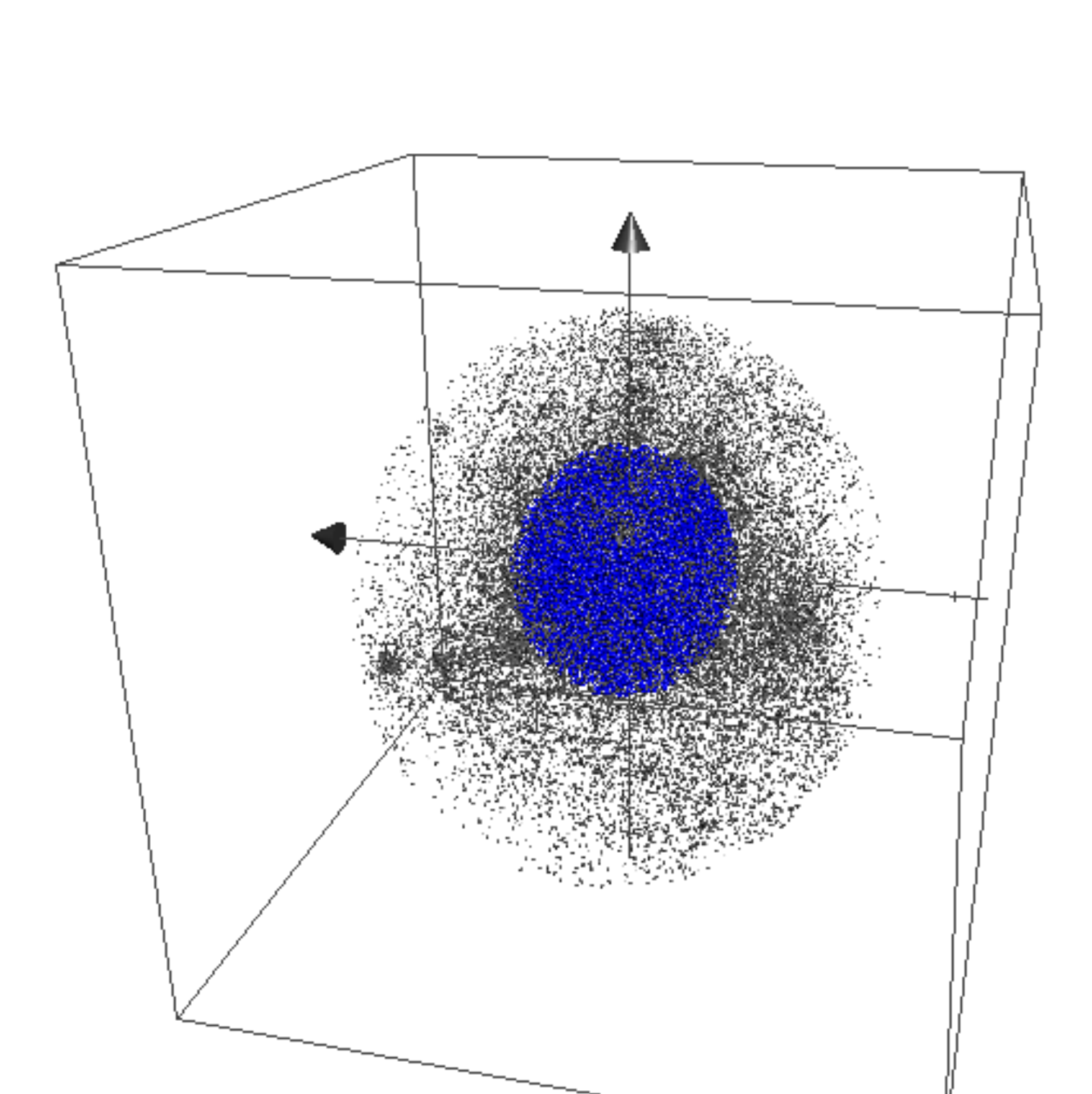}
\caption{\textit{Top panels}: spatial distribution of particles of an
  irregular halo, which is not excluded by \textit{method 1} and
  \textit{method 2} - the top-left and top-right panels show
    two different projections It has a virial mass $M_{vir}\simeq
  1.37\times 10^{14}M_{\odot}h^{-1}$ and it is clearly still in a
  merging phase, being composed by multiple mass clumps.  The black
  dots show the virial particles, while those identified at
  $500\rho_{c}$ are in blue, located around the centre of mass and the
  most massive clump.  In this case, while
  $ar1_{vir}=(a/c)_{vir}=0.21$, the axial ratio in the inner shell is
  $ar1_{500}=0.68$, causing an irregular shape profile.  This halo is
  successfully excluded by \textit{method 3}.  \textit{Bottom panels}:
  for comparison, we show the particle distribution of a regular halo,
  again with two different projections; in this case the
  axial ratios at both overdensities are approximately $0.3$.
\label{testhalo}}
\end{figure*}

\begin{table*}
\centering
\begin{tabular}{|c|c|c|c|c|c|c|}
  \hline
  name & box [Mpc $h^{-1}$] & $z_{i}$ & $m_{p}$[$M_{\odot}h^{-1}$] & soft
   [kpc $h^{-1}$] & $N_{h-tot} (z=0)$ & $N_{h>1000} (z=0)$\\
  \hline
  \textbf{Ada} & 62.5 & 124 & $1.94\times 10^{7}$ &   1.5 & 2264847 & 36561 \\
  \textbf{Bice} & 125 & 99 & $1.55 \times 10^{8}$ & 3 & 2750411 & 44883 \\
  \textbf{Cloe} &  250 & 99  & $1.24  \times 10^{9}$ & 6 & 3300880 & 54467 \\
  \textbf{Dora} &  500 & 99  & $9.92  \times 10^{9}$ & 12 & 3997898 &58237 \\
  \textbf{Emma} &  1000 & 99  & $7.94  \times 10^{10}$ & 24 &
  4739379 & 38636\\
  \textbf{Flora} &  2000 & 99 & $6.35  \times 10^{11}$ &  48 & 5046663
  &5298\\
  \hline
\end{tabular}
\caption{Main features of the simulations. The last two columns report
  the total number of haloes with more than 10 and 1000 particles
  ($N_{h}$)), at redshift $z = 0$.\label{tab_sim}}
\end{table*}

\subsection{Halo catalogues}
At each stored snapshot, we identified the dark matter haloes using
the Ellipsoidal Overdensity algorithm, as described in
\citet{despali13,despali14,despali16} and \citet{bonamigo15}.  This
algorithm identifies ellipsoidal haloes in numerical simulations: it
works similarly to the more common Spherical Overdensity criterion
\citep{lacey94,tormen04,planelles10,knebe11,knebe13}, with the
difference that the halo shape is refined using an iterative procedure
to find the best triaxial ellipsoid that follows the mass density
distribution - instead of forcing a spherical shape.  For example, at
the present time we used the virial overdensity
$\Delta_{vir}\simeq319$ as a density threshold for the main halo
catalogues \citep{eke96,bryan98}.  Then, we identified the haloes at
other four overdensity thresholds, corresponding to 200, 500, 1000 and
2000$\rho_{c}$ (as in \citet{despali16}); each run has been made
independently, so that the resulting shape and direction of each shell
is not influenced by the virial value.

\begin{figure*} 
\centering
\includegraphics[width=\hsize]{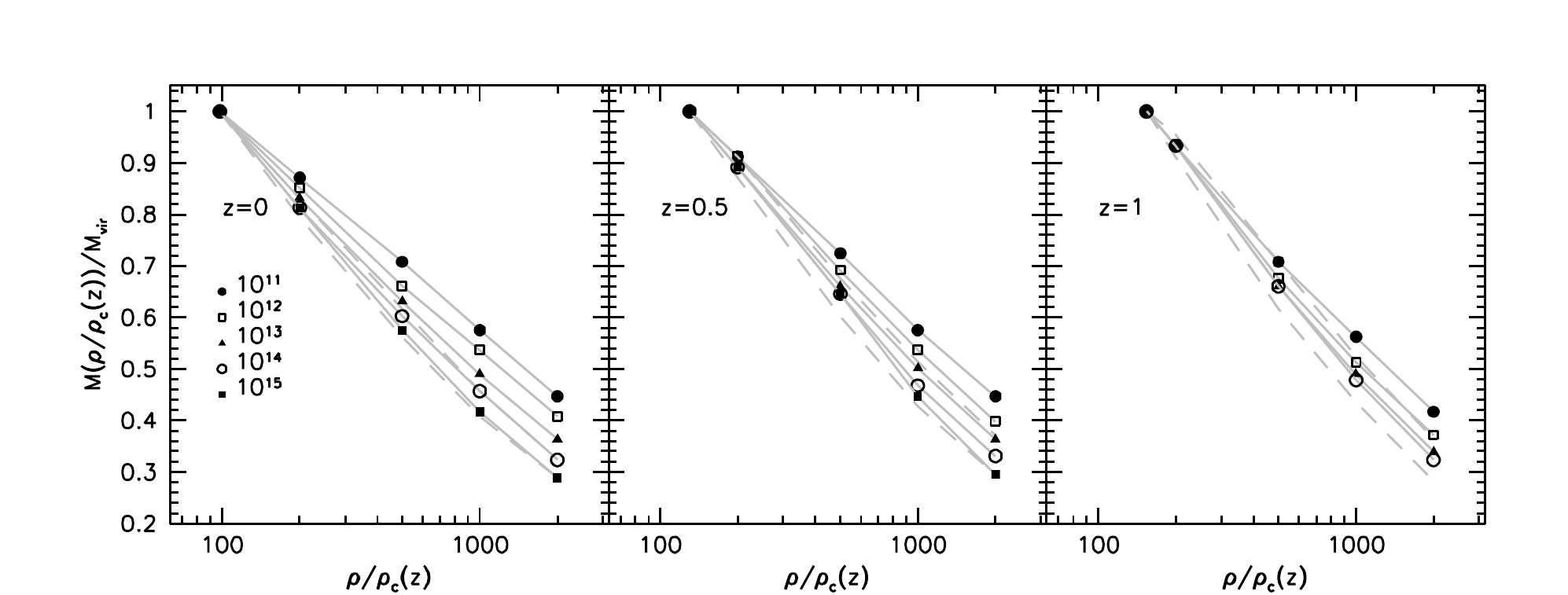}
\caption{Mass fraction in each ellipsoid, with respect to the total
  mass at the virial overdensity. Different points show the median
  result for different mass bins: in all cases the mass fraction
  decreases similarly to the centre, but with different slopes
  determined by the differences in the density profiles and thus in
  the concentration. As an example, the two dashed lines show the
  $25\%$ and $75\%$ quartiles for the mass bin associated to
  $10^{14}M_{\odot}h^{-1}$. \label{mass_rho}}
\end{figure*}

We calculate halo shapes using eigenvalues of the mass tensor, defined
as:
\begin{equation}
M_{\alpha,\beta}=\frac{1}{N}\sum_{i=1}^{N}r_{i,\alpha}r_{i,\beta},
\end{equation}
where $r_{i}$ is the position vector of the $i$-the particle and
$\alpha$ and $\beta$ are the tensor indexes.  By diagonalising
$M_{\alpha,\beta}$ we obtain the eigenvalues and eigenvector: the axes
of the ellipsoid ($a\leq b\leq c$) are then defined as the square
roots of the eigenvalues.

\begin{figure} 
\includegraphics[width=\hsize]{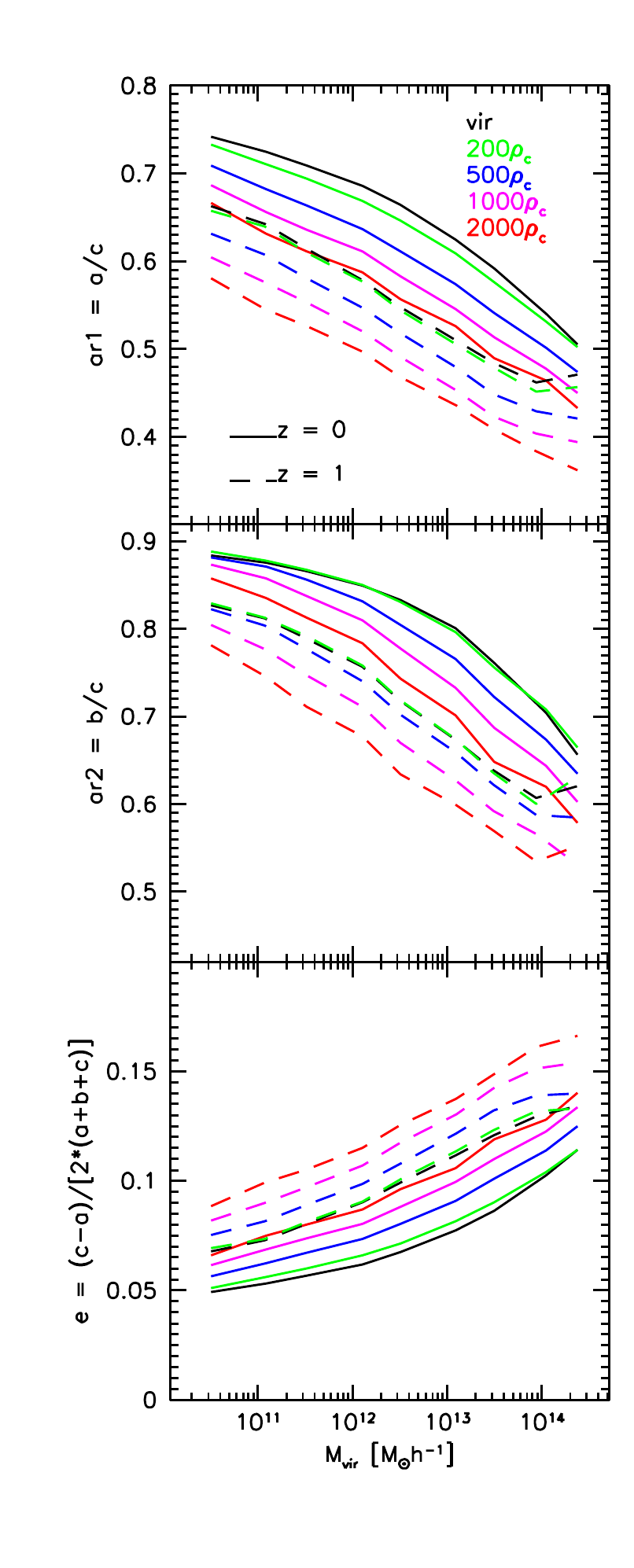}
\caption{Axial ratios and ellipticity as a function of halo mass, for
  different overdensity thresholds. The lines show the median values
  of the distributions for $ar1=a/c$, $ar2=b/c$ and
  $e=(c-a)/[2*(a+b+c)])$ with $a\leq b\leq c$).\label{shape_r_mass}.}
\end{figure}

\section{Halo selection}
Since the purpose of this work is to provide reliable prediction of
the halo shapes as a function of radius -- and overdensities, we
decided to restrict our halo catalogue in order to exclude irregular,
merging or highly unrelaxed haloes. First of all we applied one of the
common criteria to define relaxed haloes \citep{neto07,maccio08} -
\textit{method 1}: we calculated the distance between the position of
the minimum of potential and the centre of mass of the halo; then, we
maintain only systems in which this difference is less than $5\%$ of
the corresponding halo virial radius.  As seen in \citet{bonamigo15},
this criterion is able to exclude most of the irregular haloes --
meaning those that cannot be reliably described by one single triaxial
ellipsoid -- and their fraction increases with the mass, due to the
fact that high mass haloes, forming later
\citep{giocoli07,zhao09,giocoli12b}, are still in a merging phase.  As
a second criterion - \textit{method 2}, we calculated the total energy
of haloes -- as a sum of the kinetic and potential energies of the
constituting particles -- and discarded those with positive energy,
getting rid of some other irregular systems.  In Figure~\ref{sel_crit}
(top and bottom panels refer to $z=0$ and $z=1$, respectively) we
illustrate the effect of these two selection criteria on the halo
catalogue showing the percentage of irregular haloes detected (and so
excluded) by each method and by their combination (requiring
that at least one of the two methods excludes the halo).

Nevertheless,  after  this  first  selection,  we  noticed  that  some
irregular haloes were still present  in our catalogues, as for example
the two projections of the halo displayed  on top panels of Figure~\ref{testhalo}.  
This system was  chosen randomly  between  those exhibiting  extremely low  virial
axial ratios ($ar1=a/c \leq 0.2$), who  were not excluded by the first
selections:  it  is  clearly  an unrelaxed  halo,  being  composed  by
multiple mass clumps and in a  merging phase.  It survived through the
previous  selection because  (even if  it may  not be  clear from  the
figure due  to the projection  effect and the colour  combination) its
main clump is still more massive than the other, keeping the centre of
mass near the minimum of potential. From the figure, we can notice how
the shape of the $500\rho_{c}$ ellipsoid (shown in blue) is very different
from    the    overall    virial    shape    (black    dots):    while
$ar1_{vir}=(a/c)_{vir}=0.21$, the  axial ratio  in the inner  ellipsoid is
$ar1_{500}=0.68$,  in  contrast with  the  common  finding that  inner
parts are more elongated than the  outer ones (see Section 4.1).  For
comparison, the  two bottom panels,  show two projections of a  relaxed halo
with similar  mass: its axial ratios  do not change dramatically  as a
function of  radius and the  ellipsoid enclosing $500\rho_{c}$  is clearly
larger and more massive than in the previous case.

\begin{figure*} 
\includegraphics[width=0.95\hsize]{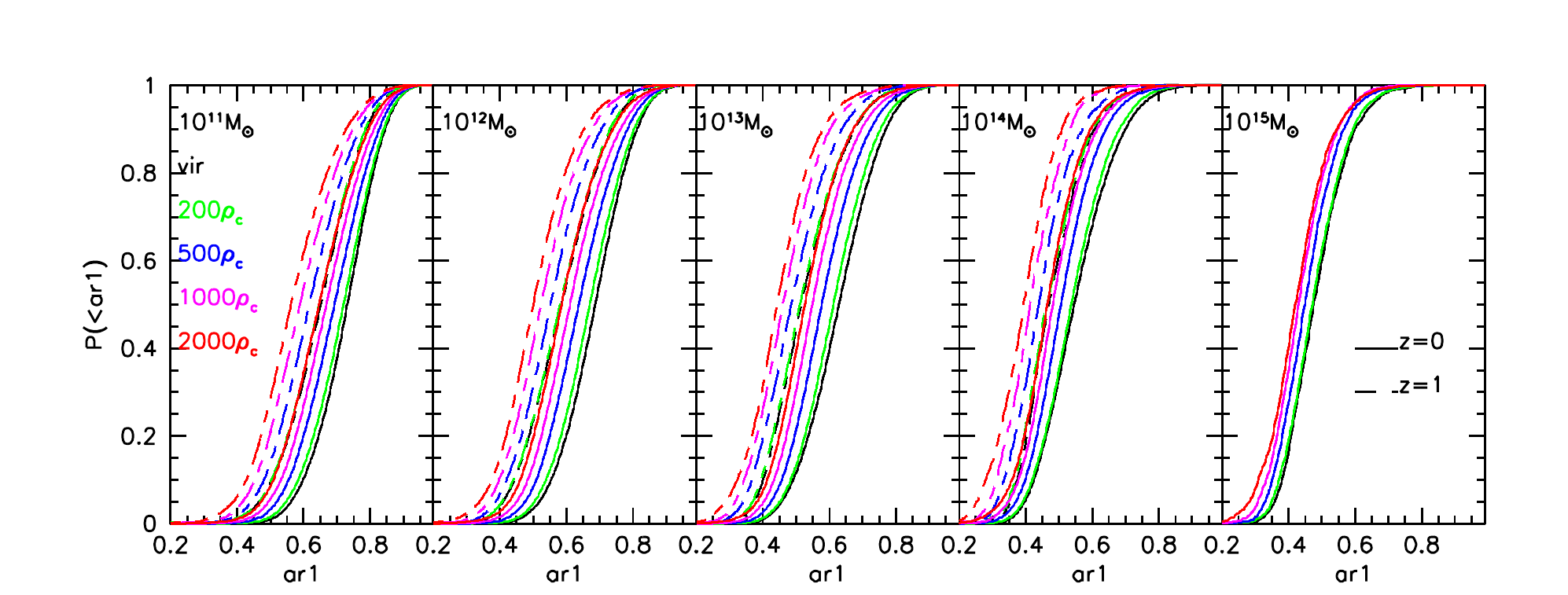}
\includegraphics[width=0.95\hsize]{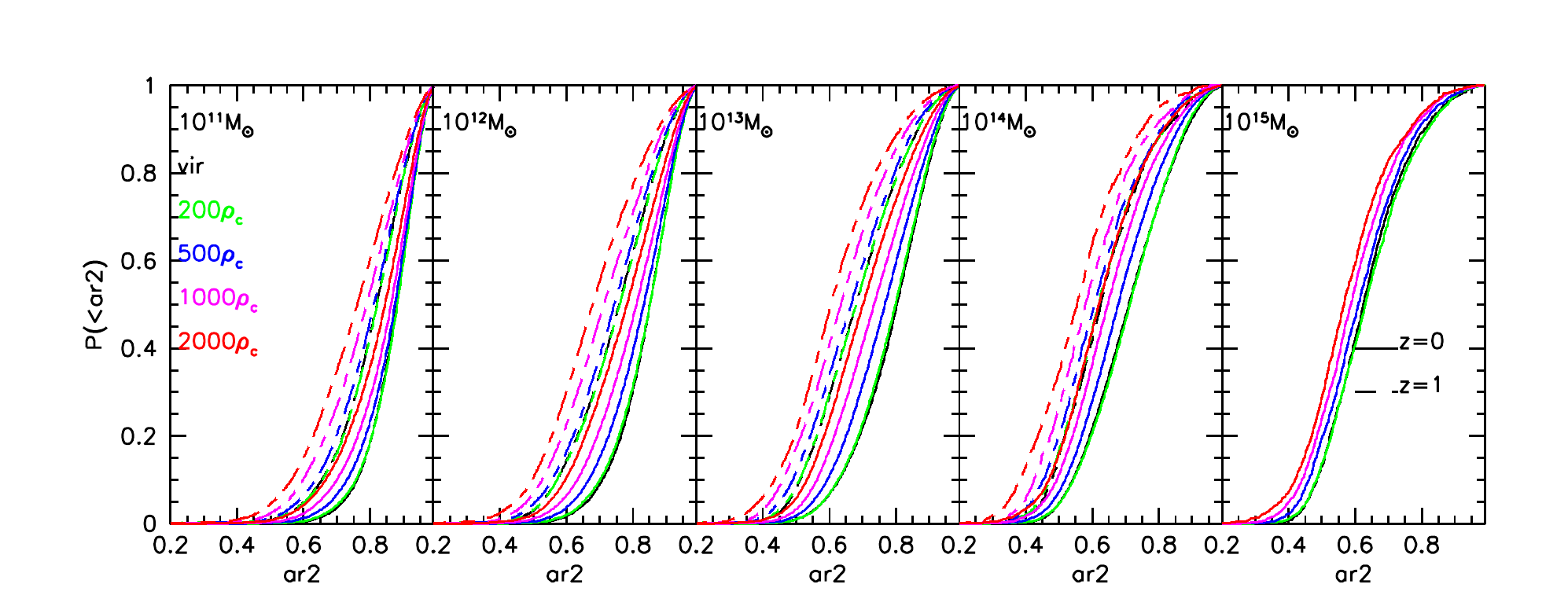}
\includegraphics[width=0.95\hsize]{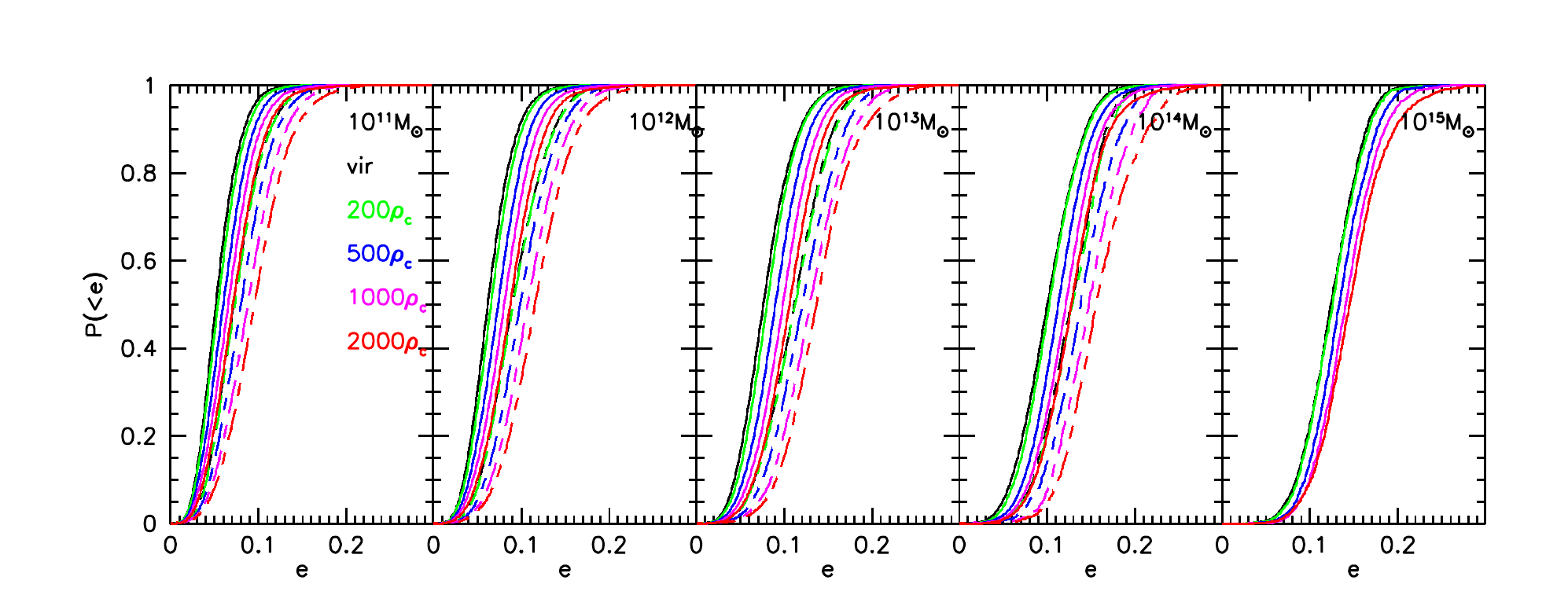}
\caption{Cumulative distributions of axial  ratios and ellipticity for
  different overdensity  thresholds; each panel shows  the results for
  haloes in a certain mass bin, centred in $10^{11} M_{\odot}h^{-1}$ -
  $10^{15}   M_{\odot}h^{-1}$,   at   redshift   0   (solid)   and   1
  (dashed).\label{3dcum}}
\end{figure*}

In order to capture systems like the one displayed in the top
  panel of the previous figure, we suggest and use a third selection
criterion (\textit{method 3}), based on the discreet measurement of
the density profile of the halo computed on elliptical shell of
defined overdensities.  In particular, we measure the mass fraction
contained in each overdensity ellipsoid, with respect to the total
virial mass.  Figure~\ref{mass_rho} shows the mass enclosed by each
overdensity threshold, as a function of the overdensity with respect
to the critical one.  The points show the median values in each mass
bin: the mass decreases with overdensity, with a mass dependent slope
because smaller haloes are more concentrated than the more massive
ones \citep{giocoli12c,meneghetti14}.  In those haloes which are in a
merging phase, the central clump can be expected to be less massive
than in relaxed haloes, since a significant part of the mass still
resides in the in-falling clumps.  Thus, this median density profile
can be used as a selection criterion, in particular to characterise a
relaxed sample in addition to the \textit{method 1} and \textit{method
  2}: we exclude also all haloes for which the mass fraction
$M_{X}/M_{vir}(\rho/\rho_{c})$ always lies in the lower quartile
($\leq 25\%$ of the distribution) for all the four discrete
overdensity ellipsoid.  We remind the reader that this last
selection criterion is very analogous to the substructure mass
fraction method adopted by \citet{neto07} and to the residuals
from a NFW fit used in \citet{maccio07}
to characterise relaxed and
unrelaxed haloes.  The fraction of haloes excluded only by
\textit{method 3} is shown by the magenta histogram in
Figure~\ref{sel_crit}: they are approximately $20\%$ of the whole halo
sample both at $z=0$ and $z=1$.  Adding \textit{method 3} as an
exclusion criterion, we are able to discard some irregular haloes that
are not identified by the first two methods: the final cut in the halo
catalogue is shown by the black histogram.  The relaxed halo in the
bottom panels of Figure~\ref{testhalo} satisfy all our selection
criteria, thus proving to be relaxed and regular -- in all considered
overdensities, while the top one is successfully excluded from the
whole sample by \textit{method 3}.

\section{3D shape as a function of overdensity}
In this section we analyse  how the three-dimensional shape of relaxed
haloes changes as a function  of the overdensity, and characterise the
variation as  a function  of mass and  redshift. This  work is
  complementary    to    what has been   done   in    \citet{despali14}    and
  \citet{despali16}; by presenting our results in terms of overdensity
  instead  of radius,  we  want to  provide  useful distributions  for
  observational studies in general and for strong lensing parametric mass modelling in
  particular, as these distributions can  be inserted as priors in algorithms
for the  representation of the position of lensed multiple image systems
  ( for example Lenstool \citep{jullo07}  ).

  We start by showing  how   the  overall   distribution  of  shapes   depends  on
  overdensity (for different masses and redshifts) and then we explore
  the conditional distribution of shapes,  binning in the virial axial
  ratio: in this way we can give realistic prediction of the change in
  shape within individual halos going from the outside to the inside.  
  Finally, we address the misalignment between the ellipsoids and show how it depends on mass and an the virial
  shape properties.

A general colour/style  code is used through the
paper: $(i)$ different overdensities are represented by various colours
(from black  for the virial case  to red for the  innermost one, going
through green, blue and magenta) and $(ii)$ $z=0$ results are shown by
solid lines, while  at $z=1$ we chose dashed lines  -- we mention that
for  better display  our data  and  results we  do not  show the  case
$z=0.5$ that lays in the middle between $z=0$ and $z=1$.

\subsection{General distributions}

First, we confirm  that halo shapes are not  self-similar as a
  function       of        distance       from        the       centre
  \citep{allgood06,vera-ciro11,jing02,bailin05}.     Figure~\ref{shape_r_mass} shows how the axial  ratios varies as a function of
the virial  mass for  our five overdensity  thresholds. We  remind the
reader that the three dimensional ellipticity is defined as
\begin{equation}
e=\frac{c-a}{2(a+b+c)}
\end{equation}
(with $a\leq b\leq c$) and it is equal to zero for a spherical system.
Apart     from    the     well     known     dependence    on     mass
\citep{allgood06,despali14,bonamigo15}, we  notice how  the dependence
on overdensity is almost the same  and regular for all relaxed masses,
with inner ellipsoids  being more triaxial than the  outermost virial one,
shown in  black. The same behaviour  can be observed at  $z=1$ (dashed
lines): at this time, for a  given mass bin, haloes are generally more
triaxial, but the inner ordering is unaltered. Note that the black and
green dashed  lines (virial and $200\rho_{c}$)  almost coincide, since
at this  redshift the two overdensities  are very close to  each other
\citep{despali16}.  The  analogous distributions for  unrelaxed haloes
at $z=0$ are shown in Appendix  A, proving why unrelaxed haloes cannot
be  easily described  by  simple  relations, and  a  regular trend  is
absent. This highlight  the fact that the  morphological properties of
galaxies and  clusters at  different radii  may be  used to  infer the
state  of  relaxation \citep{donahue16}  and  that  our relations  for
relaxed haloes do not hold for some of the recently observed clusters
(in particular for five out of six Frontier Fields clusters),
which appear unrelaxed and present multiple components.

In Figure~\ref{3dcum} we present the shape distributions in more
details: we consider five mass bins, centred on masses from $10^{11}
M_{\odot}h^{-1}$ to $10^{15} M_{\odot}h^{-1}$ and outline the
cumulative distributions of the shapes -- axial ratios and
ellipticities -- for different overdensities. From the figure we
notice that all trends are very regular
 both in mass and
overdensities.  Comparing the distributions at $z=0$ and $z=1$ we
notice that at higher redshifts haloes of same mass are more triaxial,
once the mass is fixed -- the case $z=0.5$ would lay in the middle
between $z=0$ and $z=1$ not shown here avoiding to overcrowd the
panels. As described in \citet{despali14} the redshift dependence can
be removed comparing haloes possessing the same peak-height $\nu =
\delta_c(z)/S(M)$.

\begin{table*} 
\begin{tabular}{|c|c|c|c|c|c|}    \\ \hline  
$M_{vir}[M_{\odot}h^{-1}]$ &  $\rho_{vir}$ &  $200\rho_{c}$ & $500\rho_{c}$ & $1000\rho_{c}$ & $2000\rho_{c}$ \\  
\hline 
$10^{11}$ & (0.732,0.878,0.052) & (0.720,0.882,0.054) & (0.693,0.874,0.060) & (0.669,0.863,0.066) & (0.646,0.844,0.071)\\
$10^{12}$ & (0.687,0.851,0.061) & (0.670,0.851,0.066) & (0.637,0.832,0.074) & (0.611,0.810,0.081) & (0,585,0.782,0.087)\\
$10^{13}$ & (0.629,0.803,0.076) & (0,612,0.798,0.080) & (0.577,0.767,0.090) & (0.549,0.735,0.099) & (0.506,0.701,0.106)\\
$10^{14}$ & (0.546,0.710,0.100) & (0.537,0.713,0.102) & (0.506,0.678,0.113) & (0.481,0.646,0.121) & (0.465,0.620,0.128)\\
$10^{15}$ & (0.475,0.624,0.124) & (0.471,0.631,0.126) & (0.449,0.610,0.133) & (0.430,0.587,0.141) & (0.420,0.566,0.145)\\
\hline
\end{tabular}
\caption{Median values of (ar1,ar2,e) at different overdensity $z=0$
  thresholds for five mass bins. \label{tab_shape1}}
\end{table*}

\subsection{Conditional distributions}

\begin{figure*} 
\includegraphics[width=0.95\hsize]{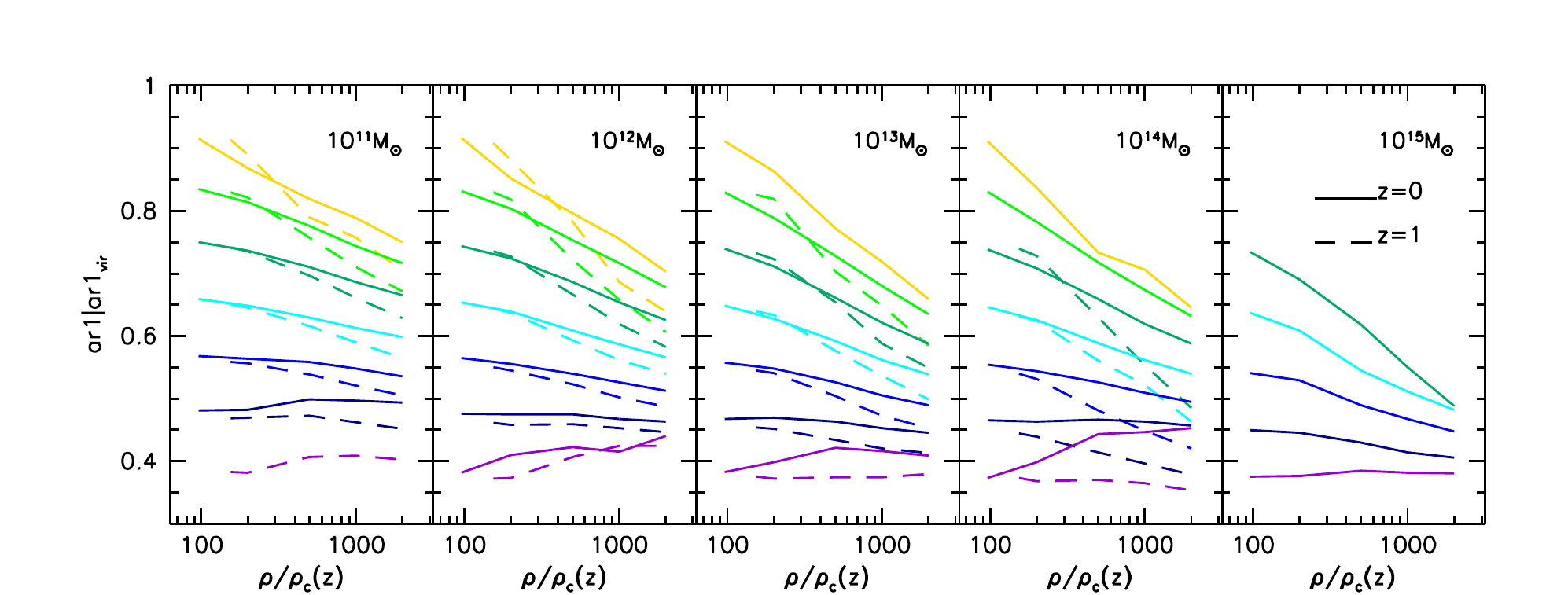}
\includegraphics[width=0.95\hsize]{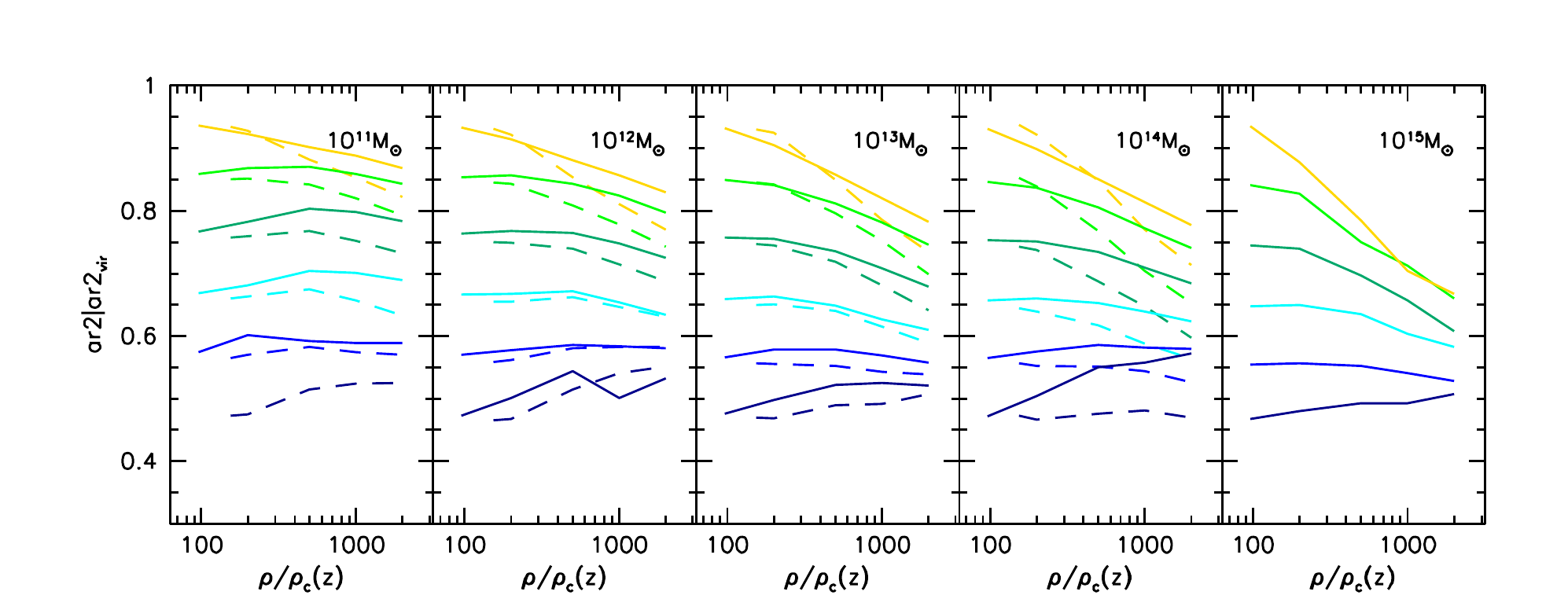}
\caption{Conditional distributions  of the axial ratios.   Each colour
  shows the median axial ratio as a function of density for the haloes
  with a  certain value of  the virial  axial ratio. For  example, the
  haloes represented by the yellow have  $0.9 < a/c \leq 1$, while for
  the purple curve $0.3 < a/c \leq 0.4$. The same holds for the second
  axial  ratio $b/c$  in the  right panel.  Solid lines  represent the
  results at $z=0$ and dashed ones those for $z=1$.\label{cond1}}
\end{figure*}

As already mentioned, strong lensing measurements focus on the
very central region  of galaxy cluster and so are  able to probe the
shape  of  the  inner  high-density   shells \citep{meneghetti16}.   At  the  same  time,
theoretical  prediction  of  mass  and   shape  at  the  virial  (or
$200\rho$) radius are often used in  their analysis and the shape in
the  central parts  is  assumed  to be  self-similar  to the  virial
one. Since we know that this is not the case and that the difference
may not be the same for all masses, we think that it is important to
link the shapes  at various distances with the virial  measurement. For this
reason, in this Section we present conditional distributions, binning
our data in virial shape.

Figure~\ref{cond1}  shows the conditional  distribution of  the
axial ratios, for  different mass bins and redshifts  $z=0$ and $z=1$,
using  solid  and  dashed  line styles,  respectively.   In  order  to
characterise  more precisely  the  shape  variation inside  individual
haloes, we bin the axial ratio distributions using the computed values
at the virial  overdensity: each colour shows  the median distribution
of  the axial  ratios for  haloes  in a  certain bin  of virial  axial
ratio. Each  bin in $ar1$  or $ar2$ has a  width of $0.1$.   Thus, the
virial  axial ratios  of haloes  represented by  the yellow  lines are
contained  in   the  interval   [$0.9$,  $1$],   the  green   ones  in
[$0.8$,$0.9$]  and  so on  and  so  forth.   The lowest  axial  ratios
$ar1/2=$[$0.3$,$0.4$] are represented in  purple colour.  Objects with
even lower axial ratios are  excluded by our selection criteria, since
extreme elongations often coincide with  haloes in a merging phase, as
already discussed  above in  the text.   A new  feature, that  was not
visible  in  the  previous   figures,  emerges  from  the  conditional
distributions: the variation of shape with overdensity -- or radius --
depends on their outer shape.  While triaxial haloes with axial ratios
$(ar1\simeq  0.5,  ar2\simeq 0.6)$  are  self-similar  in their  inner
parts,  the more  spherical ones  present a  greater shape  variation,
becoming considerably  more triaxial  inside.  This effect  is present
for all  masses, even if it  may be caused by  different phenomena: in
general, it  has been  argued that  the outer  parts of  haloes become
rounder due  to the interactions  with the surrounding  density field,
which take place after their formation, while the inner parts maintain
the original triaxiality due to the collapse process.  This is true in
particular  for low  mass haloes,  which formed  earlier and  are more
influenced by the  surrounding tidal field or by  encounters with more
massive structures.   For high-mass  haloes, apart  from the  few with
high formation redshifts, the  physical explanation of this dependence
may be  different: it is  possible that, while  matter is still  in an
accretion phase  from many directions onto  a $10^{15}M_{\odot}h^{-1}$
halo  - as  they live  at the  intersection of  filaments, the  centre
already collapsed in a well defined triaxial object.

\begin{figure*} 
\includegraphics[width=0.43\hsize]{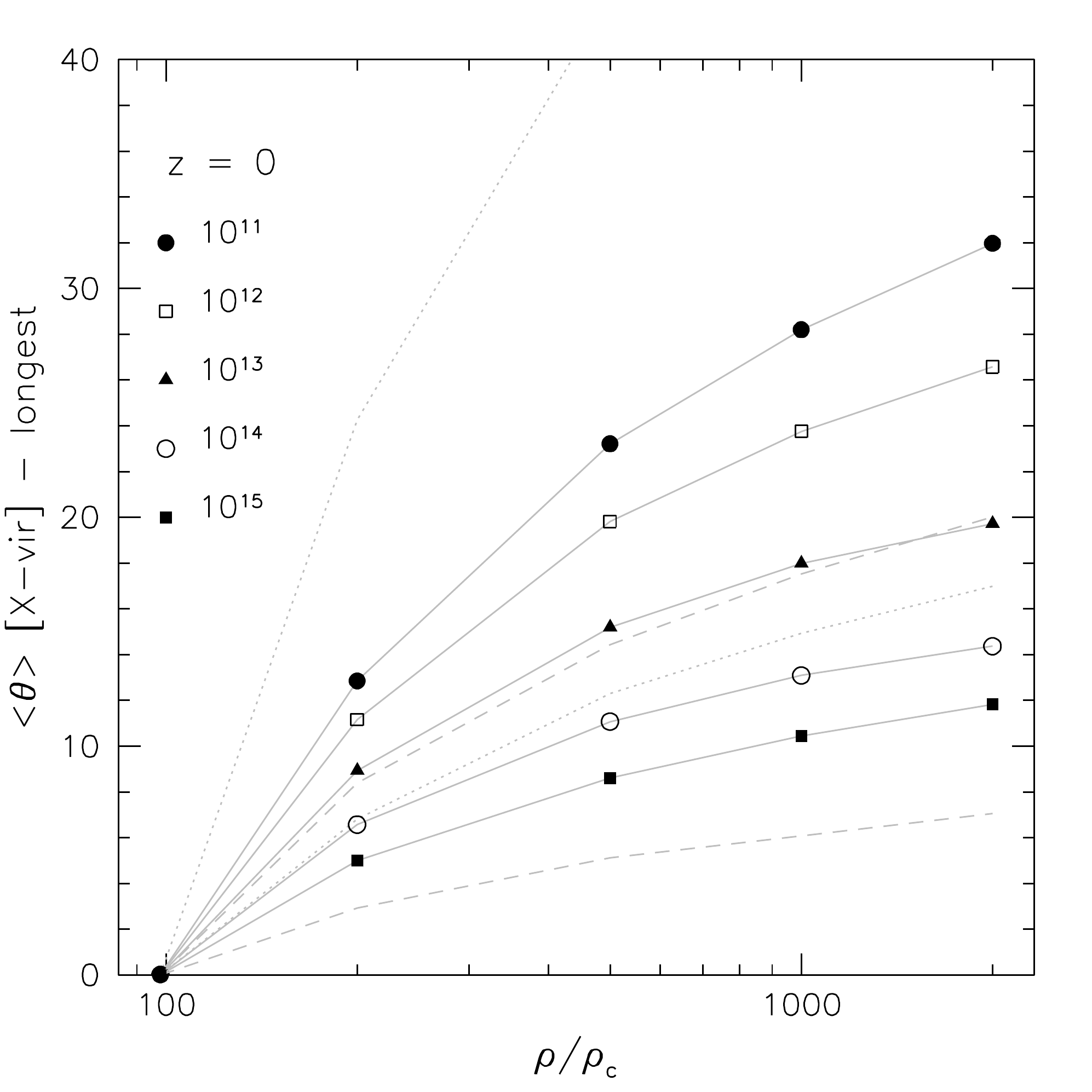}
\includegraphics[width=0.43\hsize]{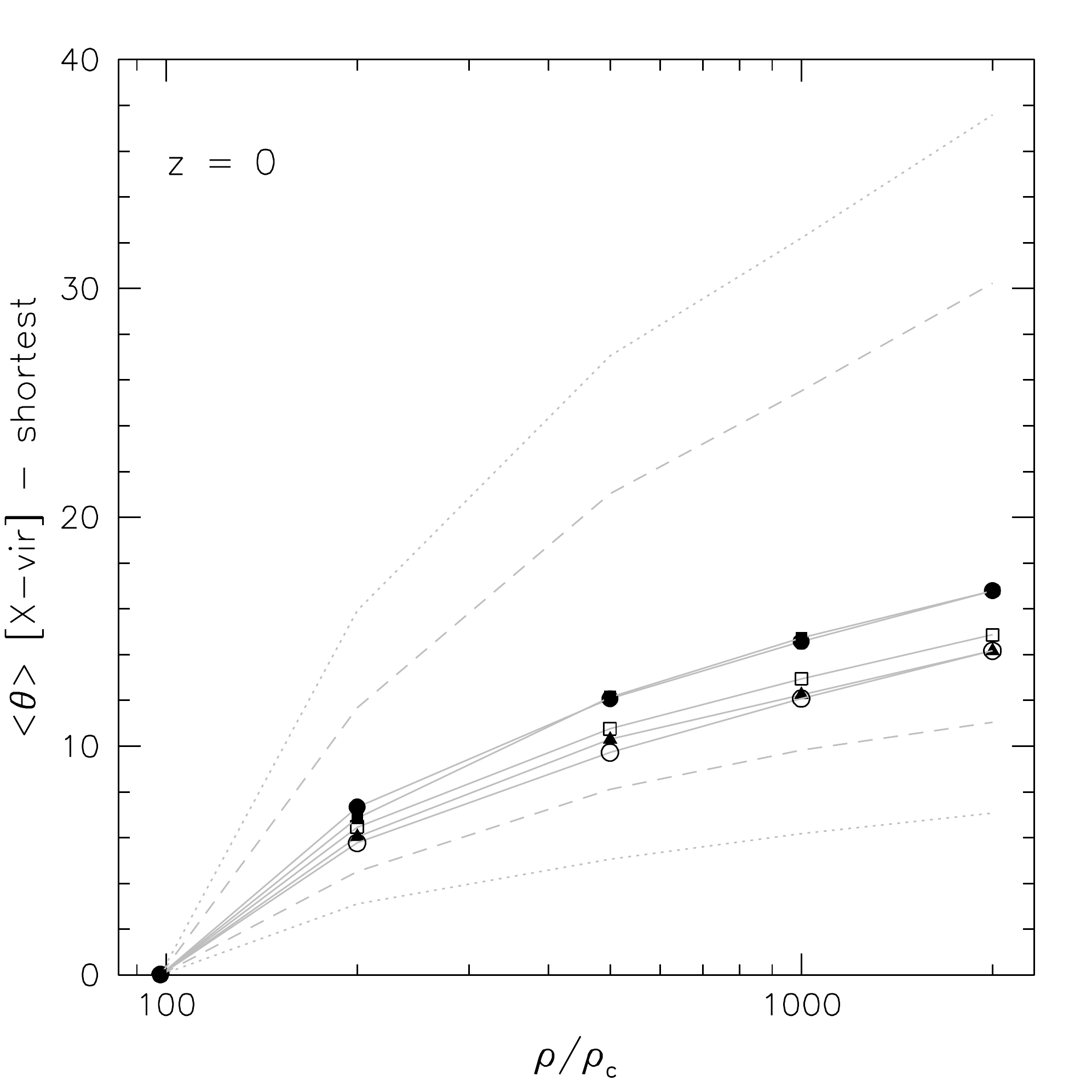}\\
\includegraphics[width=0.43\hsize]{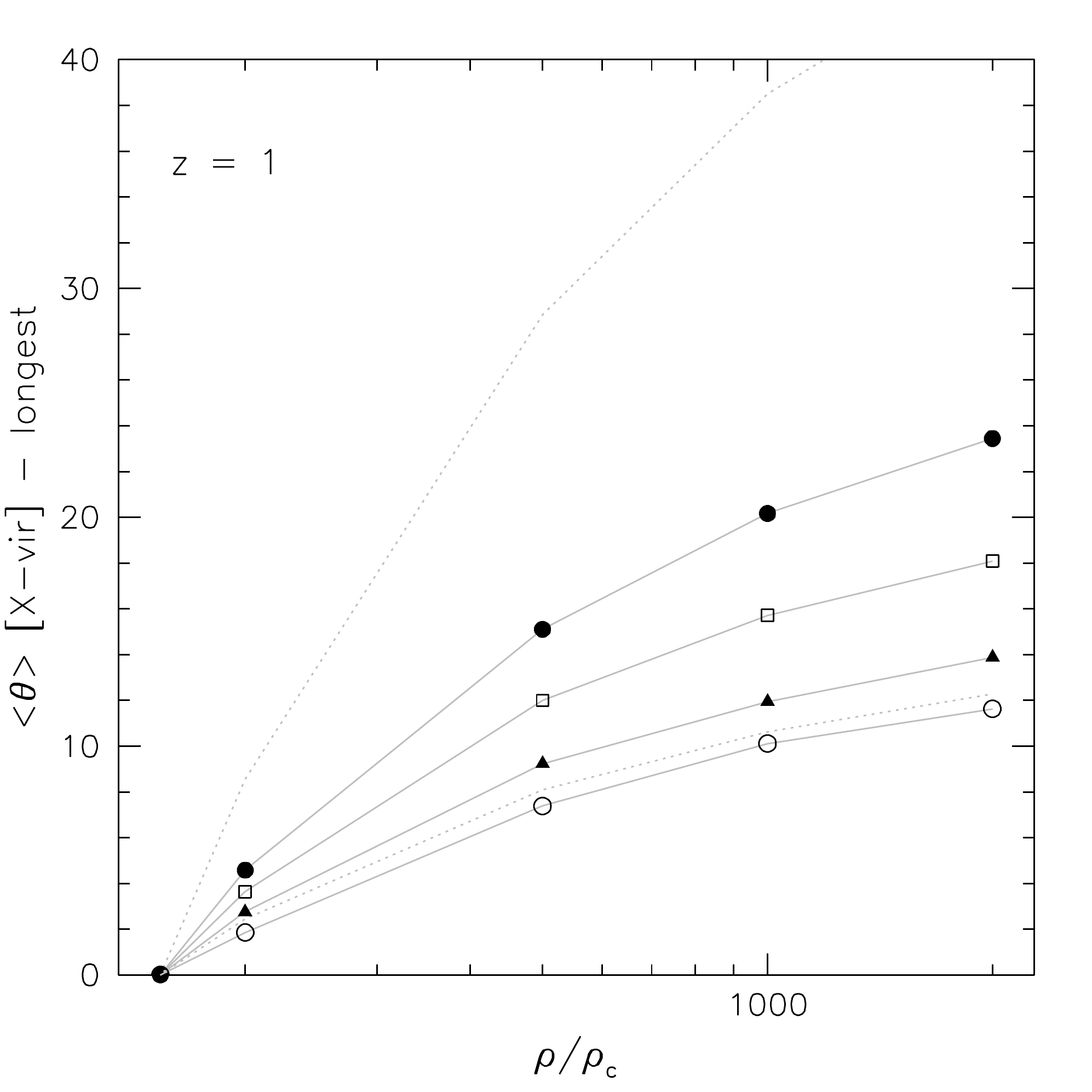}
\includegraphics[width=0.43\hsize]{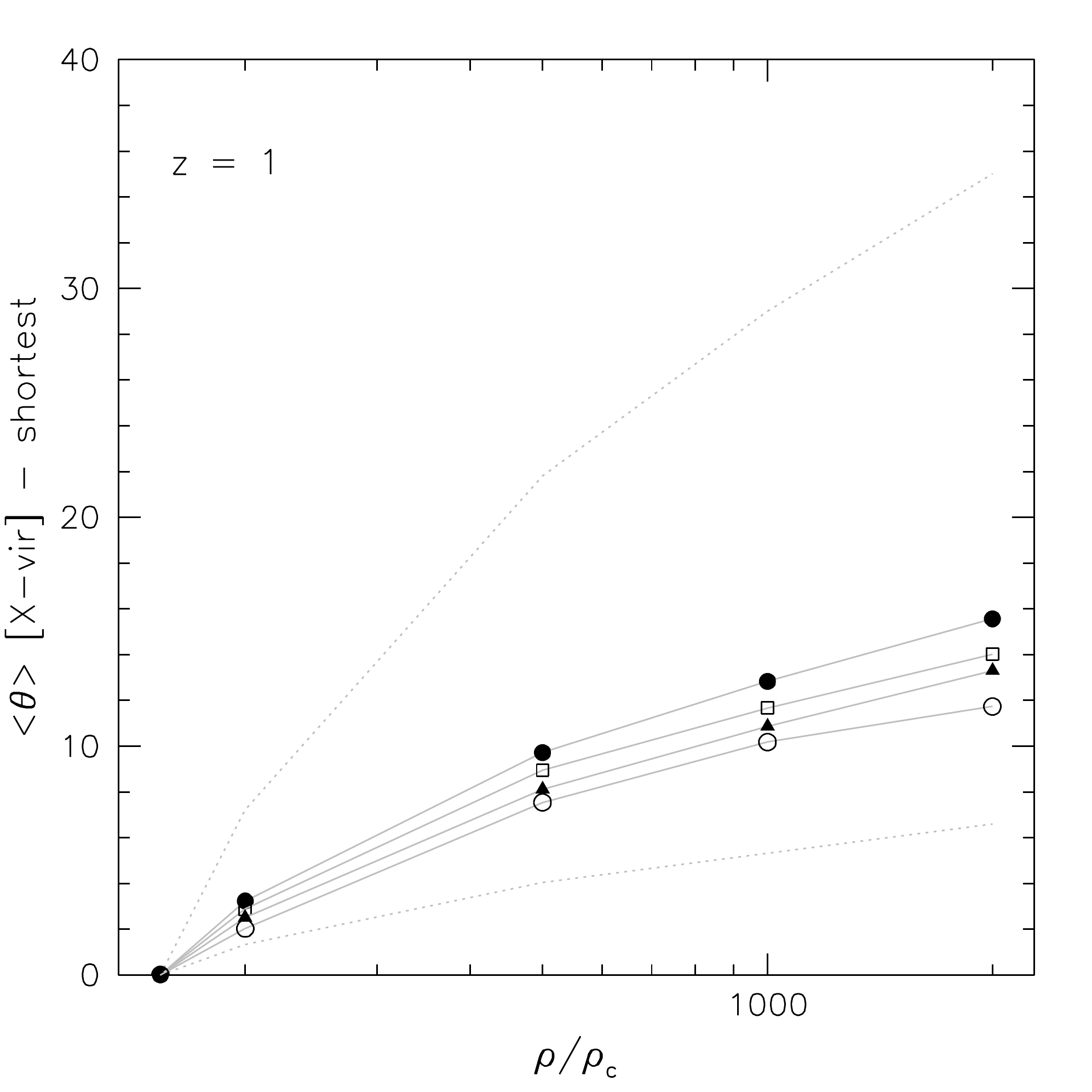}
\caption{Misalignment angle of  the four inner ellipsoids  with respect to
  the virial  one, as a function  of halo mass. The  upper panel shows
  the the median  angle between the two longest axes,  while the lower
  one the median misalignment between the two shortest axes. Different
  mass bins are  represented by different point types;  the dashed and
  the  dotted  lines   show  the  25\%  and  75\%   quartiles  of  the
  distribution         for        $10^{15}M_{\odot}h^{-1}$         and
  $10^{11}M_{\odot}h^{-1}$, respectively.\label{angle1}}
\end{figure*}

\begin{figure*} 
\includegraphics[width=0.43\hsize]{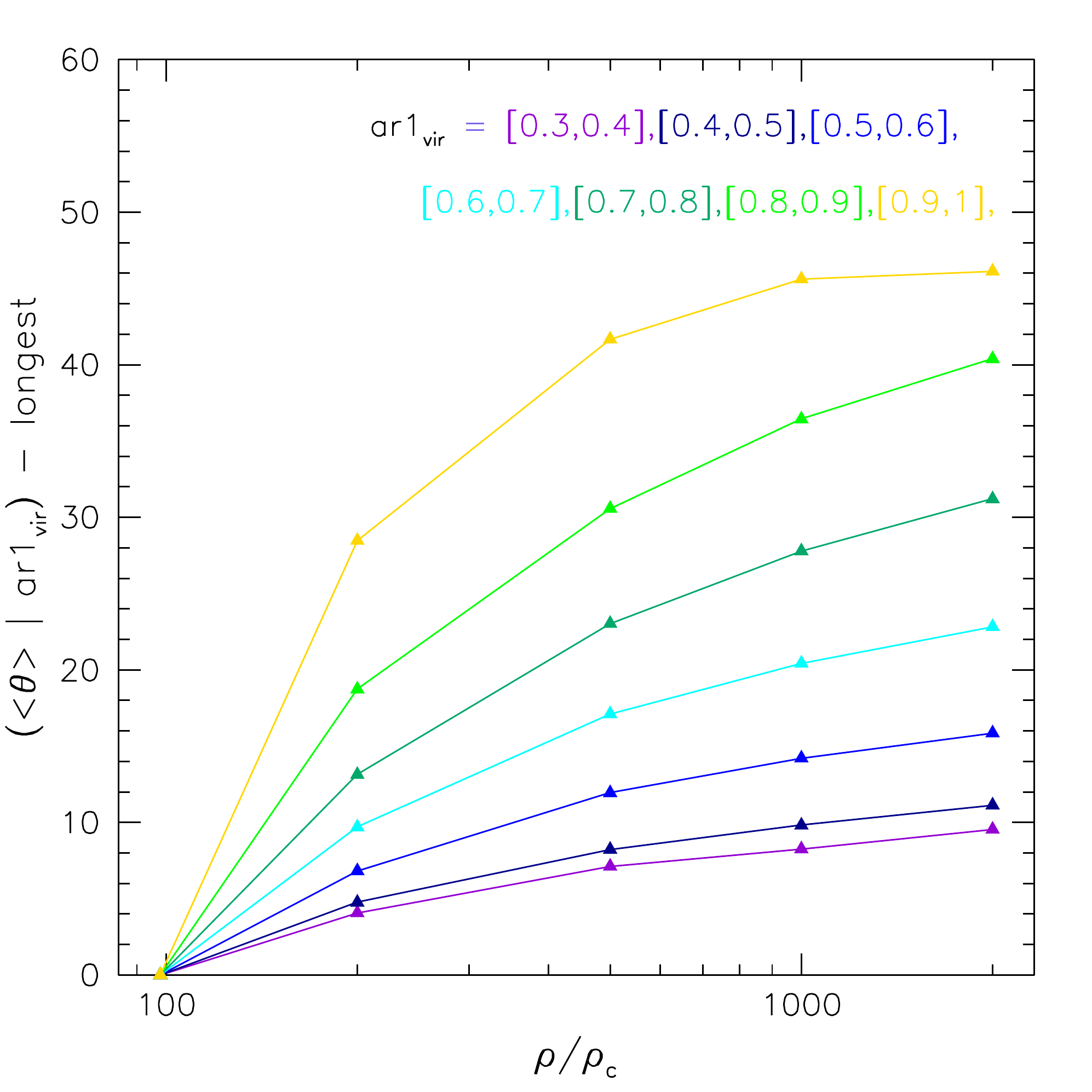}
\includegraphics[width=0.43\hsize]{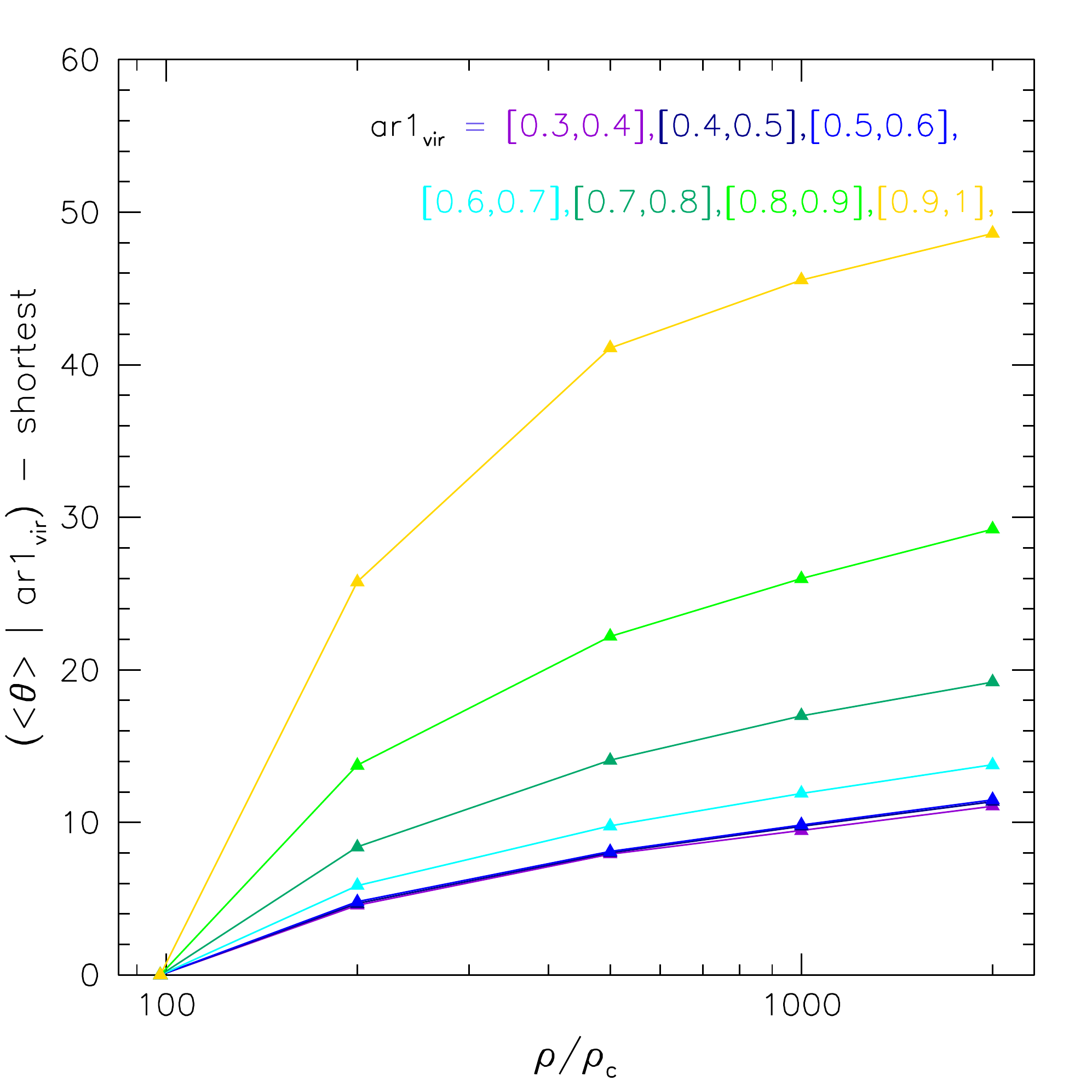}
\caption{Conditional misalignment of the different
  ellipsoids. Different colours show angles corresponding to haloes
  with a certain $ar1_{vir}$, as was done for the axial ratios in
  Figure~\ref{cond1}. Since binned results where almost the same for
  the five mass bins of the previous figure, we decided to not bin in
  mass in this case: this proves how the scatter of
  Figure~\ref{angle1} between different halo masses is due to the
  different distribution of shapes.\label{angle2}}
\end{figure*}

We underline that these distributions are also useful to generate mock
mass density distributions of dark matter in galaxies and clusters
producing for example more realistic lens models \citep{giocoli12a}.
Moreover, using the virial shape as a prior in the analysis of
strong lensing clusters and assuming that the halo shape is
self-similar at all radii can introduce a bias in the calculation,
as this is true only for a subset of objects and the innermost part
of the halo can be much more elongated than the outside
\citep{meneghetti16}. Looking at Figure~\ref{cond1}, we also notice
that the axial ratio of the central parts of the haloes tend to
converge to similar values in all cases: this means that retrieving
the true virial shape from observations, using only the inner shape
as an information, can be risky. For this reason, the shape
informations must be combined with those on mass, concentration and
other halo properties in order to derive meaningful estimates.

\subsection{Misalignment of the different ellipsoids}

\begin{figure*} 
\includegraphics[width=0.45\hsize]{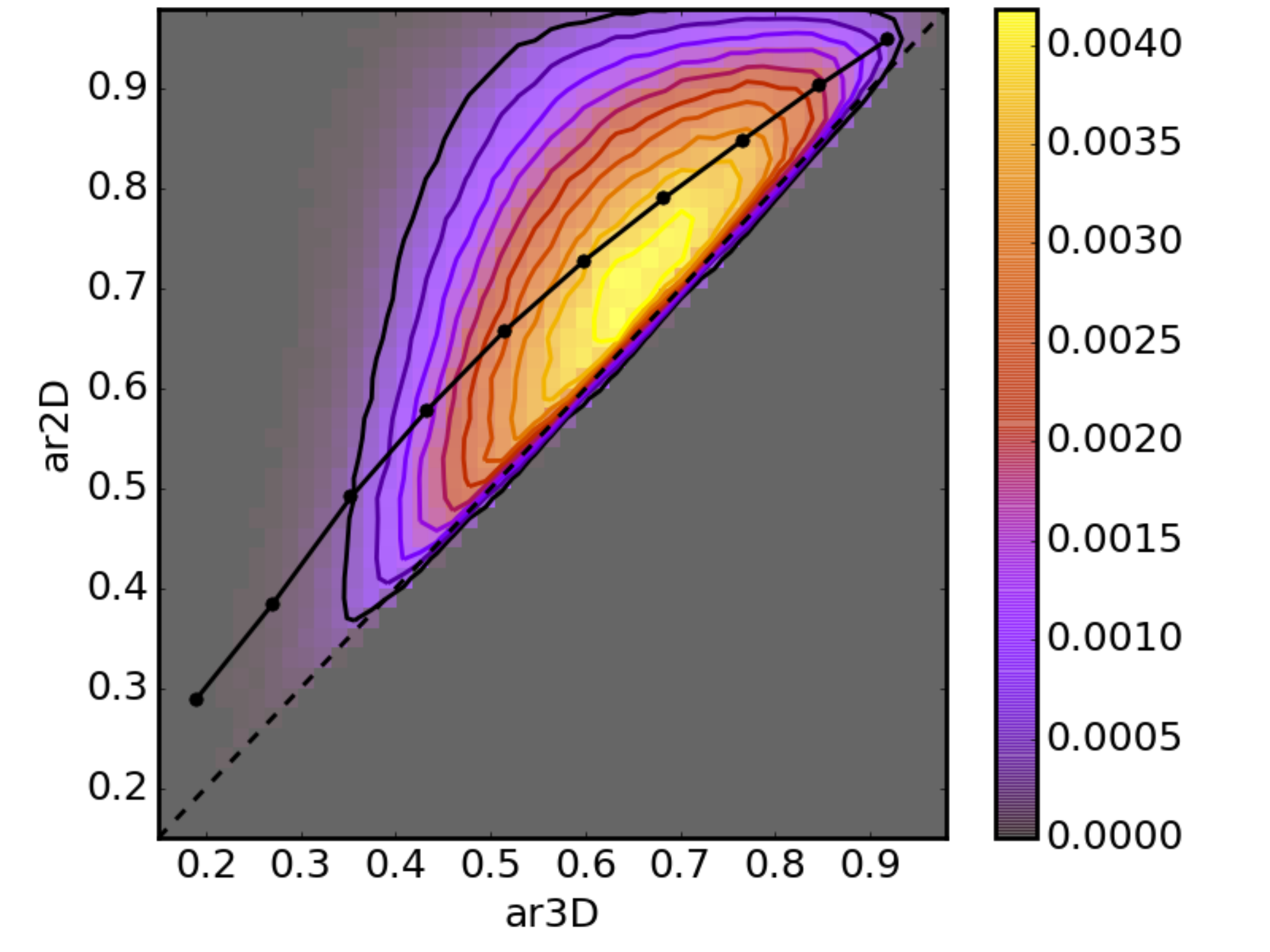}
\includegraphics[width=0.45\hsize]{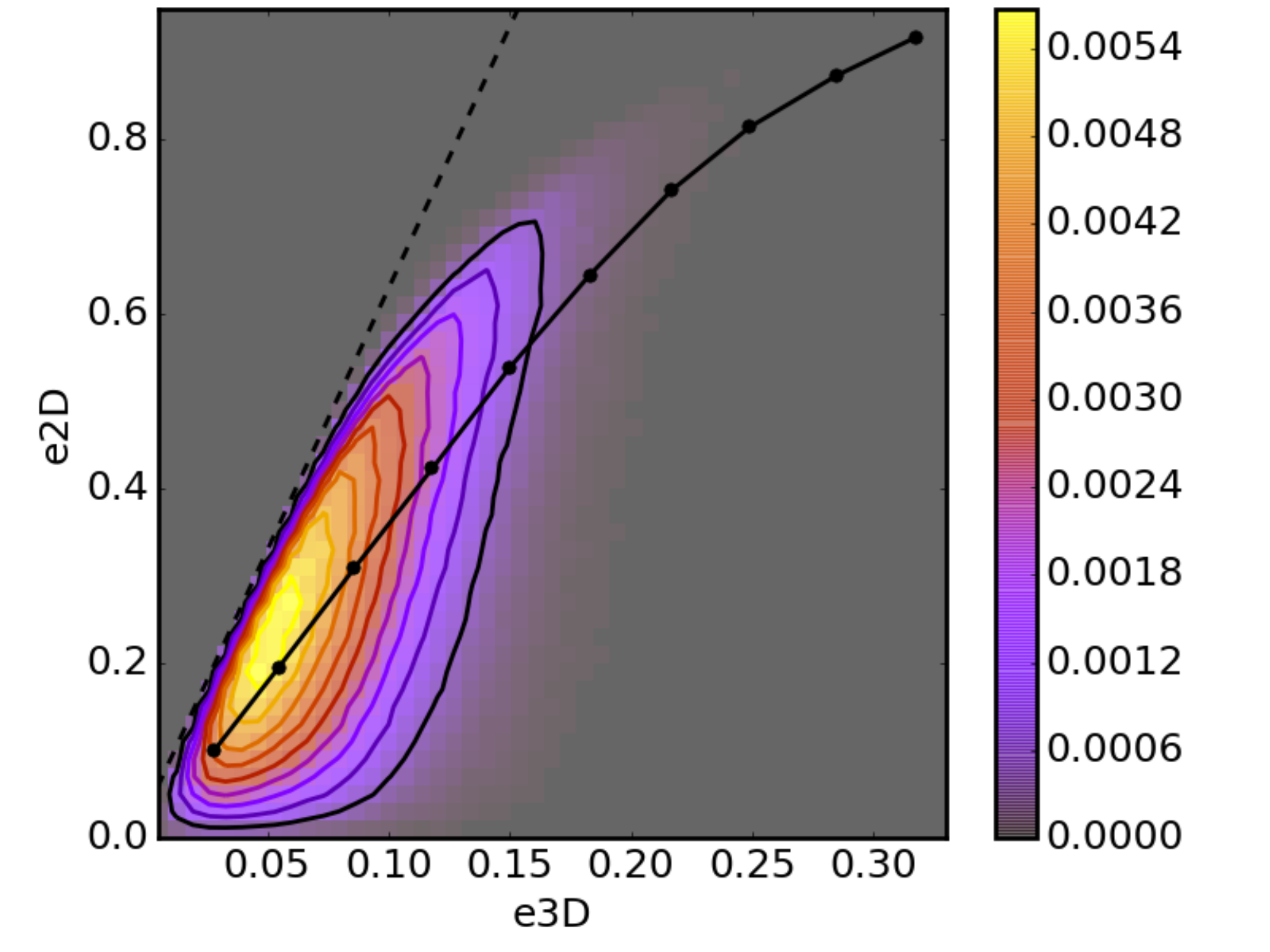}
\caption{Correlation between the 3D and 2D shapes. For each real halo,
  we calculated three two dimensional projections and here we plot all
  of  them. In  both cases,  the axes  have been  calculated from  the
  inertia tensor,  in three  or two  dimension, but  the ellipticities
  follow  two   different  formula:  $(c-a)/(2*(a+b+c))$  in   3D  and
  $(a^{2}-b^{2})/(a^{2}+b^{2})$   for  2D.   In   both  cases,   $e=0$
  corresponds to a sphere.  We used  the first axial ratio and so, for
  simplicity  in  this  particular  case,  ar3D  is  what  was  called
  previously ar1.  Since the projection  is just a geometrical effect,
  here  we   show  the  points   corresponding  to  all   the  density
  thresholds.\label{2d3d}}
\end{figure*}

It has been previously shown that shapes measured at different radii
within the same halo are not perfectly aligned with each other. Having
independently measured the triaxial shapes of four inner ellipsoids,
we can easily compare their relative orientations -- with respect to a
predefined direction of the three dimensional ellipsoid -- in a way to
better understand how on average the different ellipsoids are
misaligned with each other.  We remind the reader that our shape
measurements include all the components of haloes and do not
discriminate between the main smooth component and the substructures,
as has been done in other previous works \citep[as for
  example][]{vera-ciro11}.  In Figure~\ref{angle1} we show the median
misalignment angle between the four inner ellipsoids with respect to
the virial one -- that we view as reference; we considered the
misalignment between the two longest ($left$) and the two shortest
axes ($right$) of the 3D mass ellipsoid -- this will give us
also an idea of the deformation of the triaxial mass ellipsoid.  The
measurements are divided in five mass bins, represented by different
point types.  From the figure we notice that while the median
misalignment at $200\rho_{c}$ is around 10 degrees only, it becomes
larger when going toward the halo centre.  We stress that there is a
considerable dispersion in the data, which increases for low mass
haloes: to give an idea about this, the dashed (dotted) lines show the
25\% and 75\% quartiles associated with the highest (lowest) mass bin
- thus $10^{15}$ ($10^{11}$).  The median misalignment appears to
depend on mass: in particular, for cluster size haloes, which formed
very recently -- or are in their formation phase, the various
ellipsoids are aligned with each other within 10 degrees - probably to
the direction of compression of the gravitational collapse, while low
mass haloes -- that formed typically at higher redshifts
\citep{lacey93,giocoli07} -- present greater variations, again due to
the interactions with the surrounding field and their evolution after
the formation time (more evident is the halo to halo variation) --
also being less gravitationally strong they tend to be more influenced
by the surrounding density field.  As shown in \citet{despali14}, the
ellipticity anti-correlates with the formation redshift and so more
elongated haloes formed more recently.  Also, this mass dependence may
hide a geometrical shape dependence: low mass halos are rounder and
thus the axes direction at the virial radius may be less defined than
in a very elongated system.  This seems to be supported also by the
results of Figure~\ref{angle2}, where we display the conditional
distributions of the misalignment of the different ellipsoids
enclosing various overdensities. The data are divided according to the
virial axial ratio $ar1$, as in Figure~\ref{cond1}.  From this figure
it appears more clear that for very triaxial haloes all the ellipsoids
are well aligned, probably due to the phase of collapse or a recent
formation, while this is not true for rounder haloes.  This figure
shows all the masses together, as we noticed that the conditional
distribution are very similar for different mass bins, reinforcing our
framework and our interpretation.

Together with the results on conditional axial ratios presented at the
beginning of this  section, these distribution can be  used to produce
self-consistent mock  mass density distribution of  realistic triaxial
and perturbed haloes.

\section{Projected 2D shape as a function of overdensity}

After the  discussion presented  in the  previous section  about three
dimensional shapes, we proceed analysing  the shapes in two dimensions
(2D),  which can  be more  directly related  with observed  quantities
projected  on the  plane of  the sky  and that  have not  been
modelled in previous  works. We present the 2D  results with figures
similar  to  the  3D  ones   discussed  in  the  previous  sections:
cumulative distributions  for different overdensities and  mass bins
and conditional distribution of the axial ratios.

We project each halo along three random directions in particular along
the  three   axes  of   the  coordinate   system  of   the  simulation
($x$,$y$,$z$) -- considering  each projection a random  measure of the
2D shape  of the halo ellipsoid.   Since we already have  a relatively
large number  of haloes we do  not consider necessary to  project each
object along different  possible random line of sights.   We then look
at the  distributions of  halo shapes  and orientations  for different
overdensities and masses, as done and discussed for the 3D case. In 2D
we calculate the ellipticity as
\begin{equation}
e =\frac{a^{2} - b^{2}}{a^{2} + b^{2}} ,
\end{equation}

As a general result, 2D-distributions maintain the same properties and
ordering of the 3D ones, but they become shallower due to projection
effects. Even the extreme cases, such as very elongated shapes, are
blurred by being projected in random directions.  This general effect
can be seen in Figure~\ref{2d3d} where the contours show the point
density of the relationship between the axial ratios (left) and
ellipticities (right) as measured in 2D versus the 3D ones.  The black
points show the median of the distribution at a fixed 3D value.

\begin{figure}
  \begin{center}
\includegraphics[width=0.85\hsize]{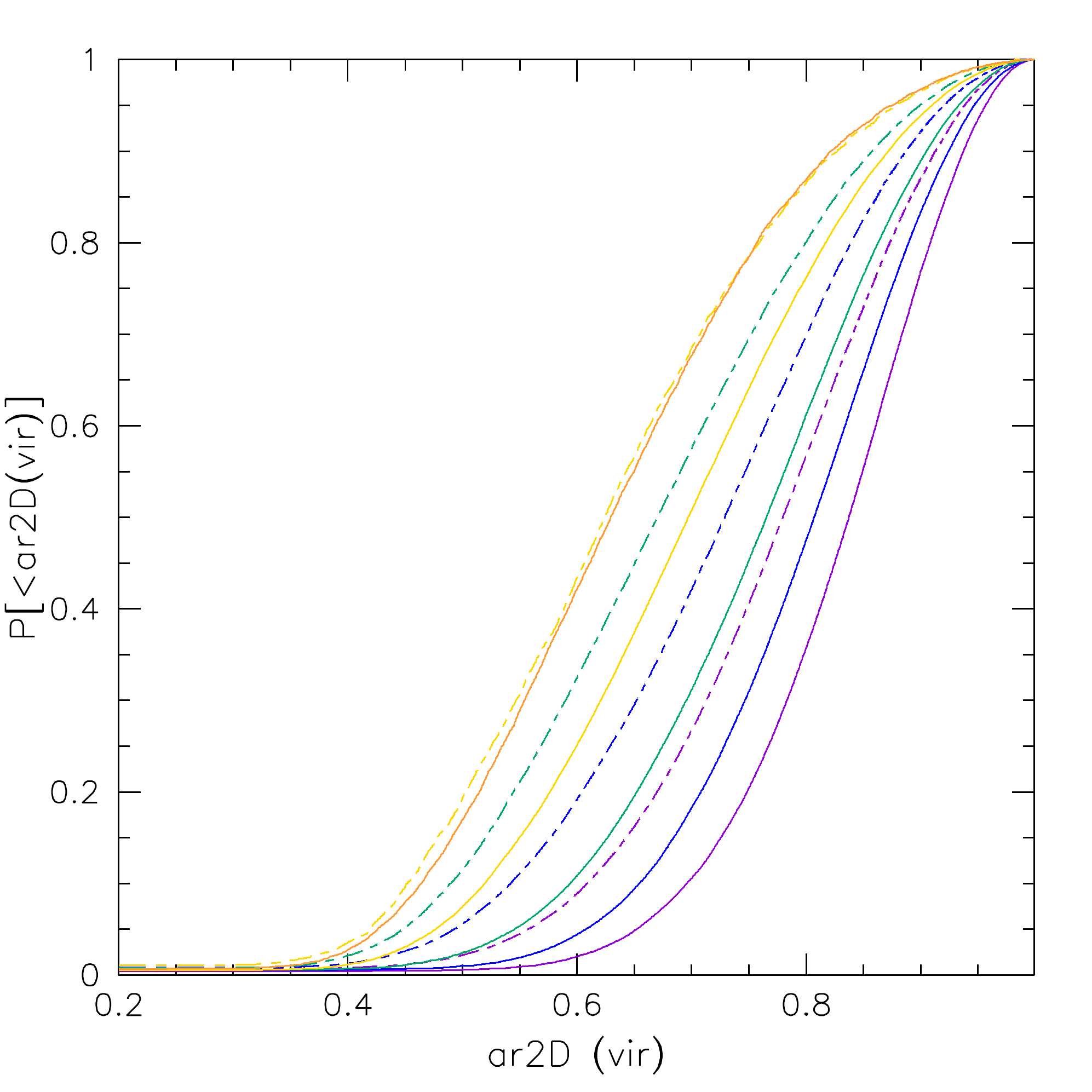}
\includegraphics[width=0.85\hsize]{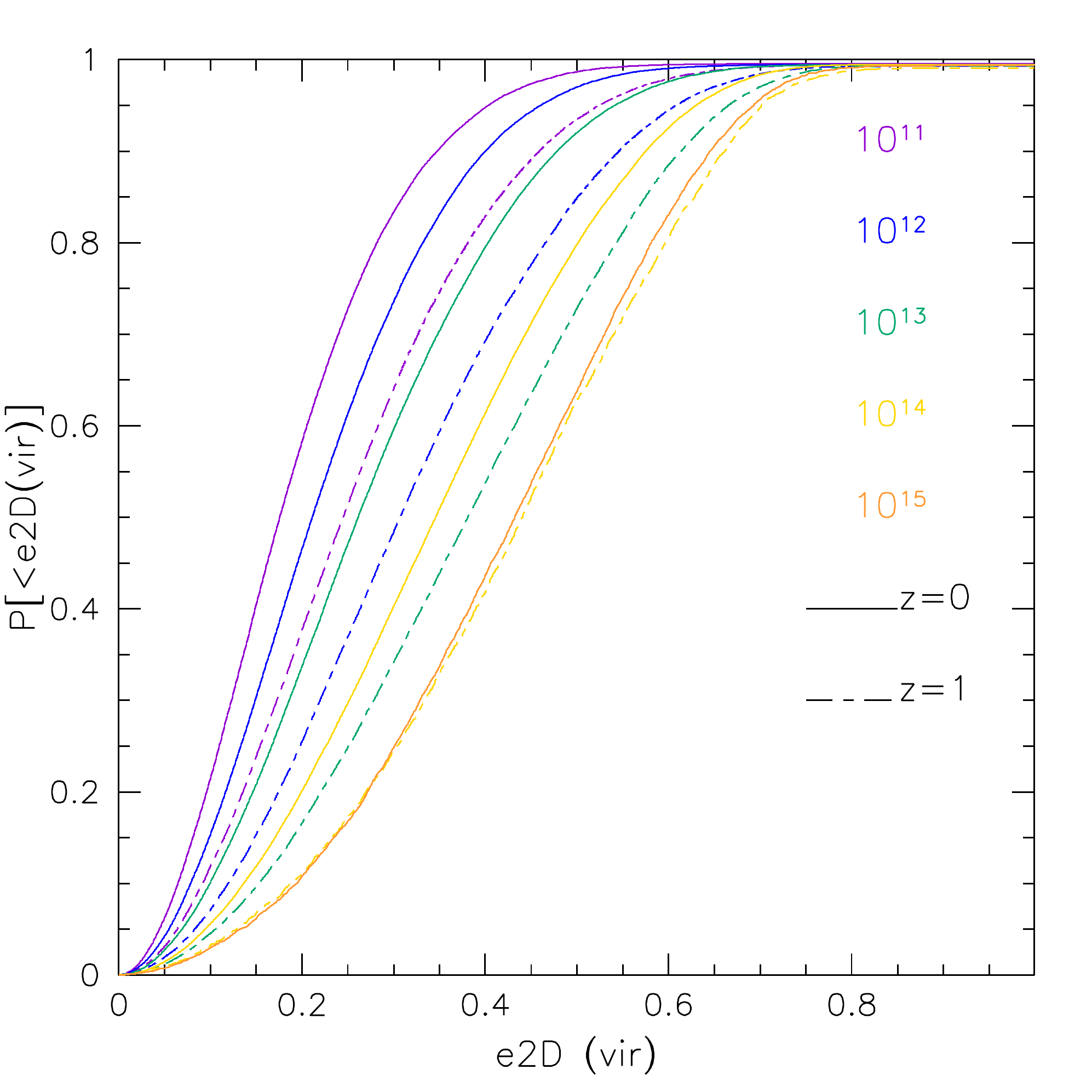}
\end{center}
\caption{Projected axial ratios and ellipticities  in 2D at the virial
  overdensity. The  different colours  show various halo  masses while
  solid  and  dashed  curves  refer   to  redshift  $z=0$  and  $z=1$,
  respectively\label{2dmass}}
\end{figure}

\begin{figure*} 
\includegraphics[width=0.95\hsize]{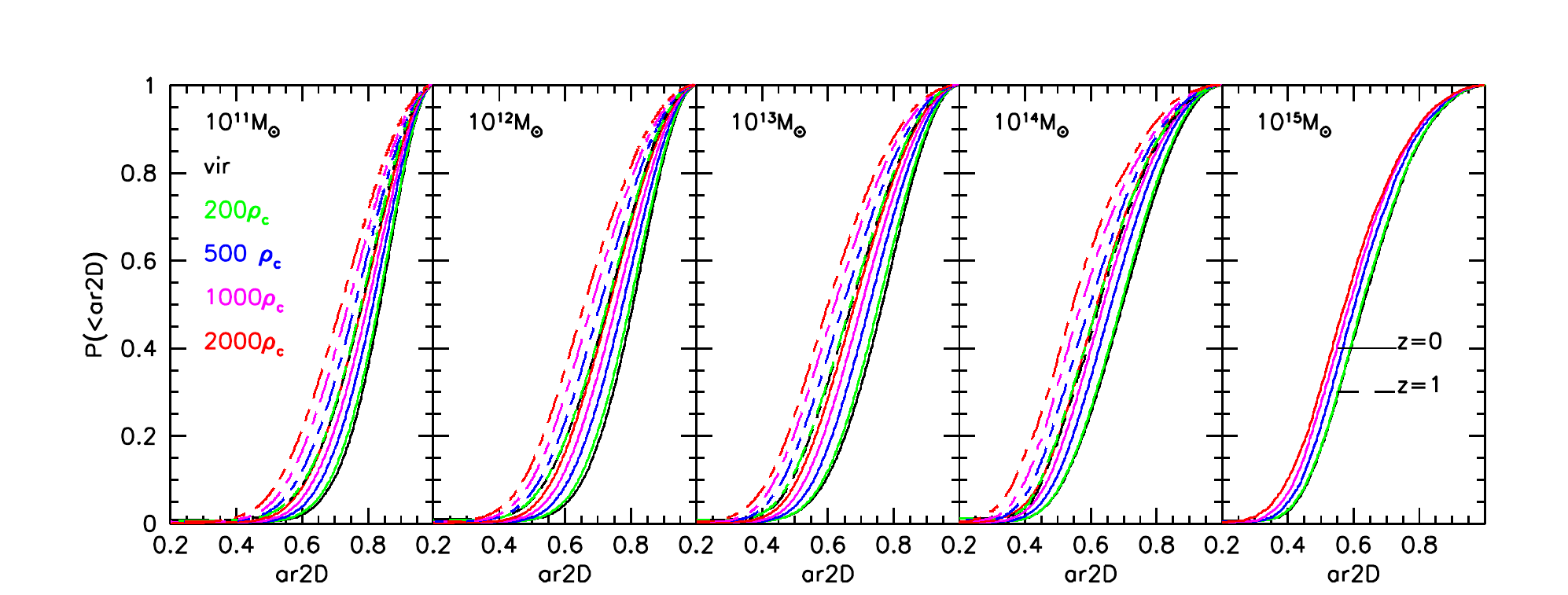}
\includegraphics[width=0.95\hsize]{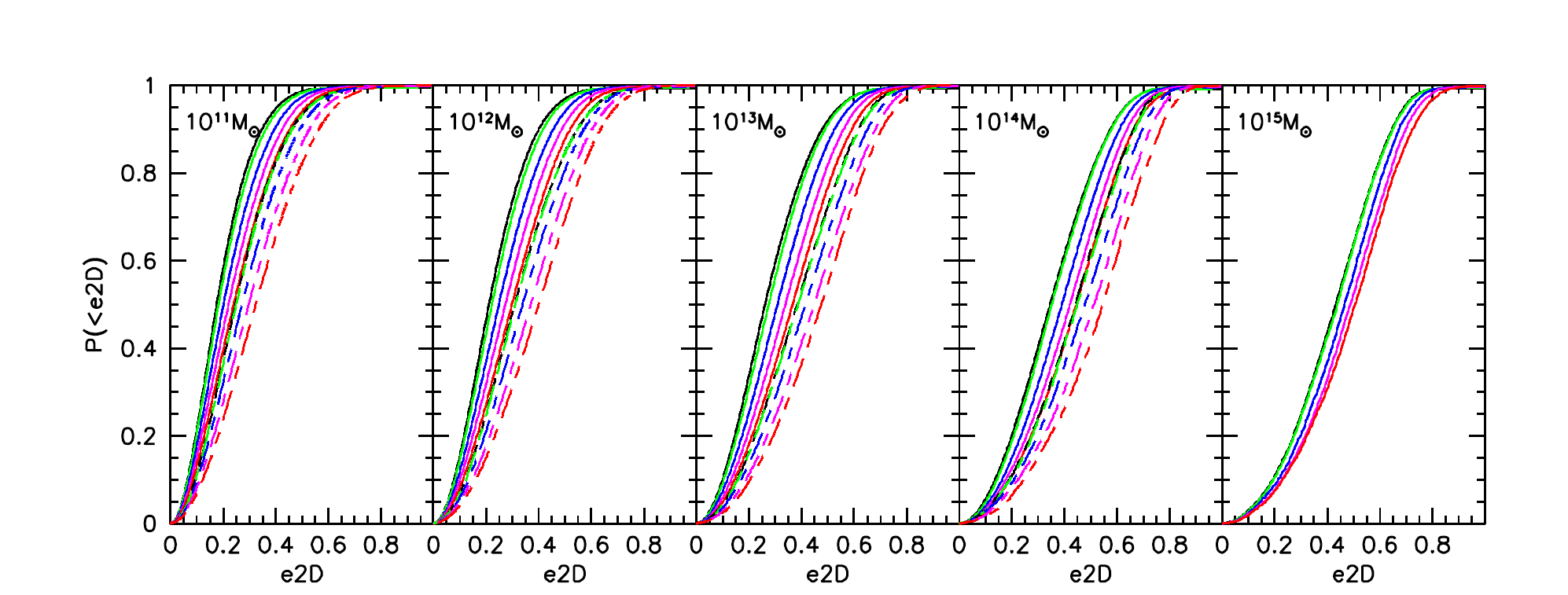}
\caption{Cumulative  distributions  of   projected  axial  ratios  and
  ellipticities. The  panel from  left to right  show the  results for
  various  host  halo masses  while  the  colours refer  to  different
  overdensities:  virial in  black, green  $200$, blue  $500$, magenta
  $1000$ and red $2000\rho_c$.  Solid  and dashed line styles refer to
  redshift $z=0$ and $z=1$ respectively.
  \label{2drho}}
\end{figure*}

In Figure~\ref{2dmass} we present the cumulative probability
distribution of axial ratios (top panel) and ellipticities (bottom
panel) for haloes of different masses and at redshift $z=0$ (solid
curves) and $z=1$ (dashed curves).
The ordering due to the different overdensity definitions is not
altered when one looks at two dimensional quantities
(Figure~\ref{2drho}), even if the low-axial ratios tail is reduced in
comparison with the three dimensional quantities, as presented in
Figure~\ref{3dcum}.

\begin{figure*} 
\includegraphics[width=0.95\hsize]{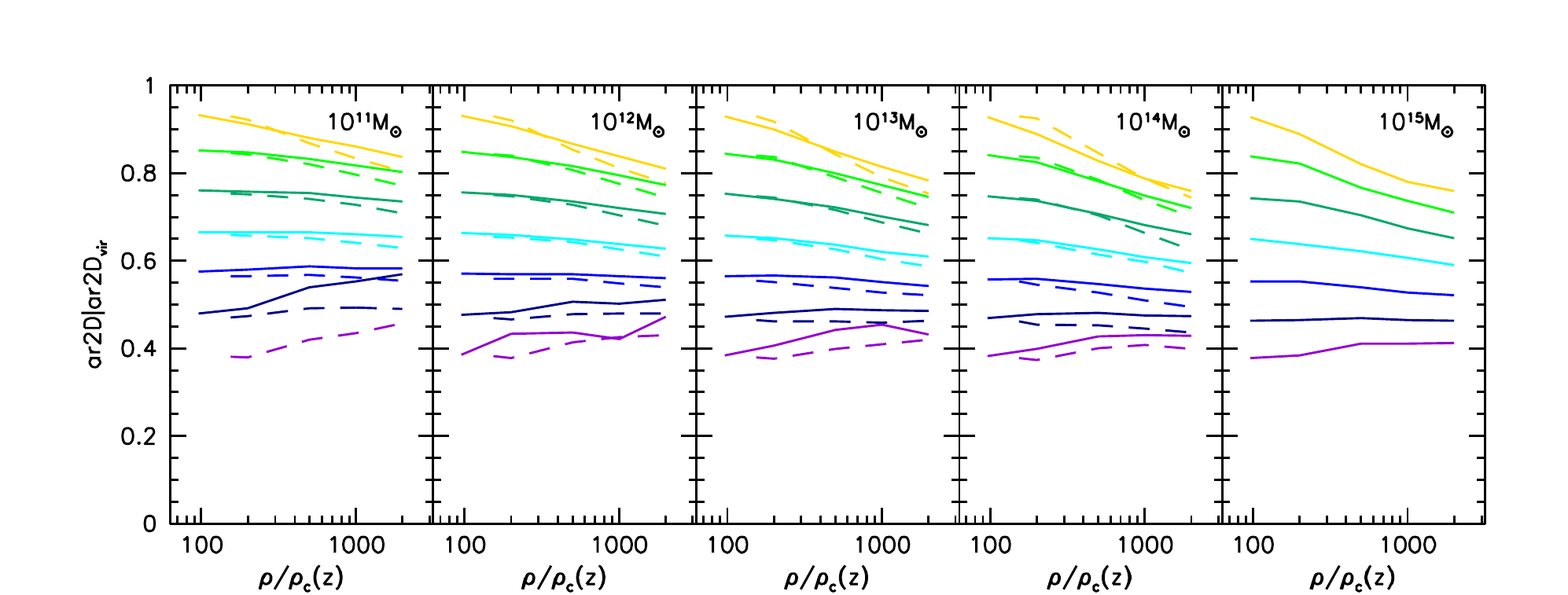}
\includegraphics[width=0.95\hsize]{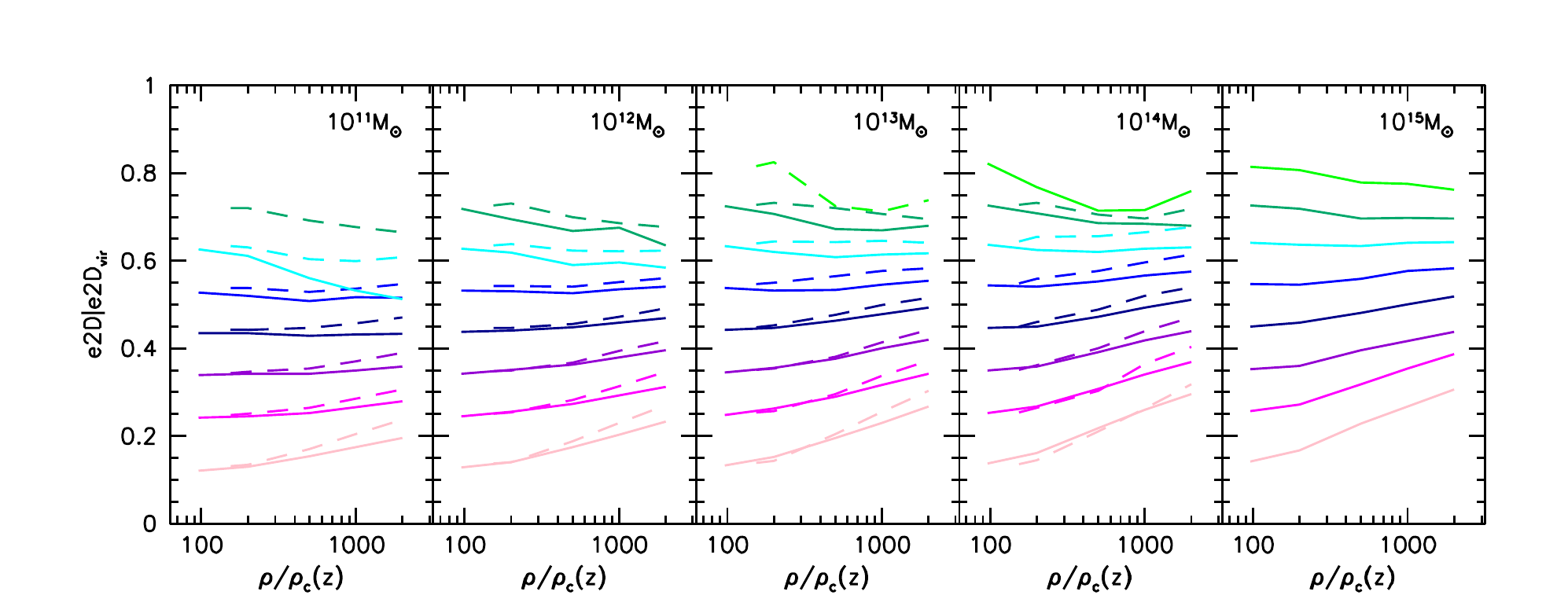}
\caption{Conditional distributions  of the  projected 2D  axial ratios
  and ellipticities.  Each colour  shows the median  axial ratio  as a
  function  of density  for the  haloes with  a certain  value of  the
  virial axial ratio. \label{2dcond}}
\end{figure*}

\begin{figure*} 
\includegraphics[width=0.45\hsize]{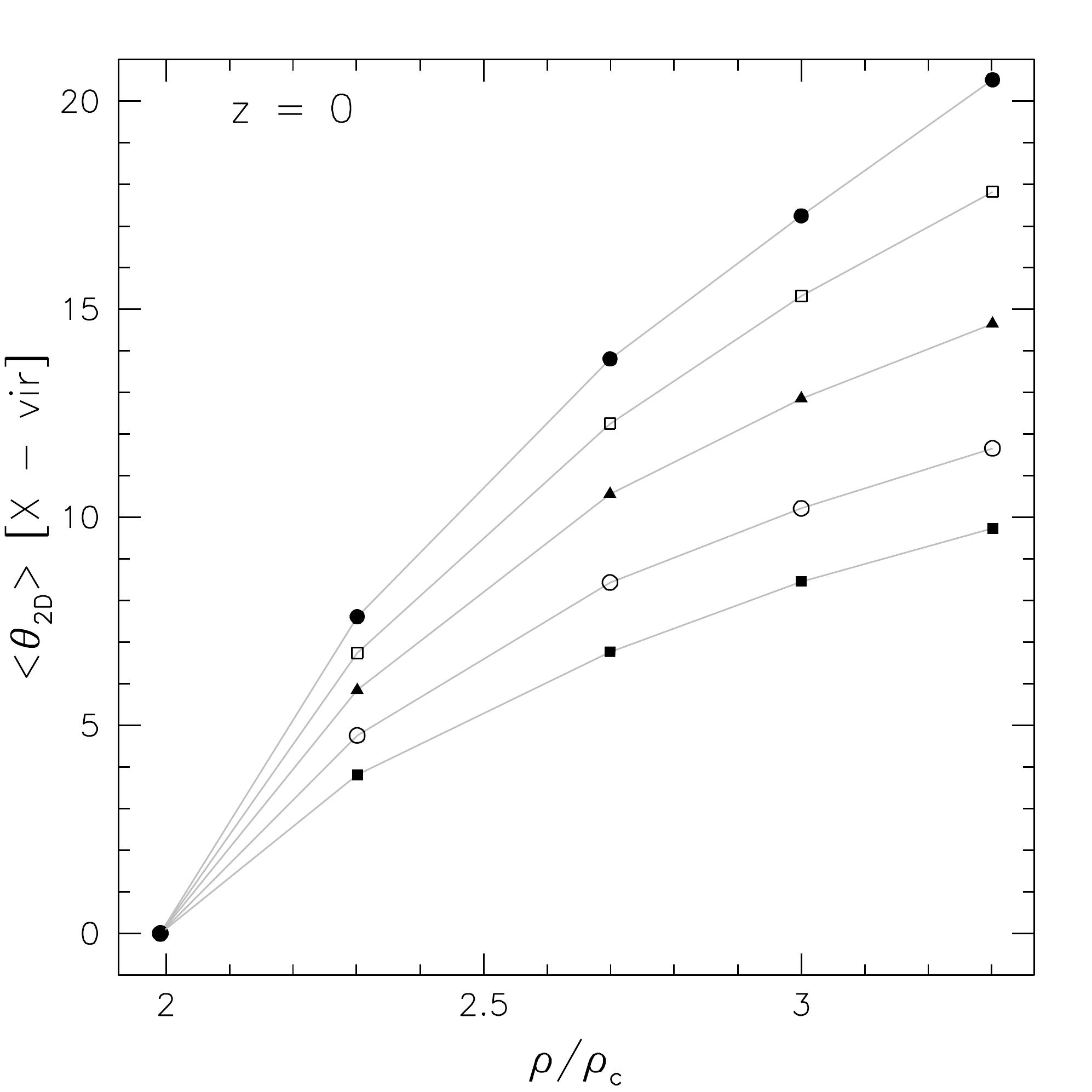}
\includegraphics[width=0.45\hsize]{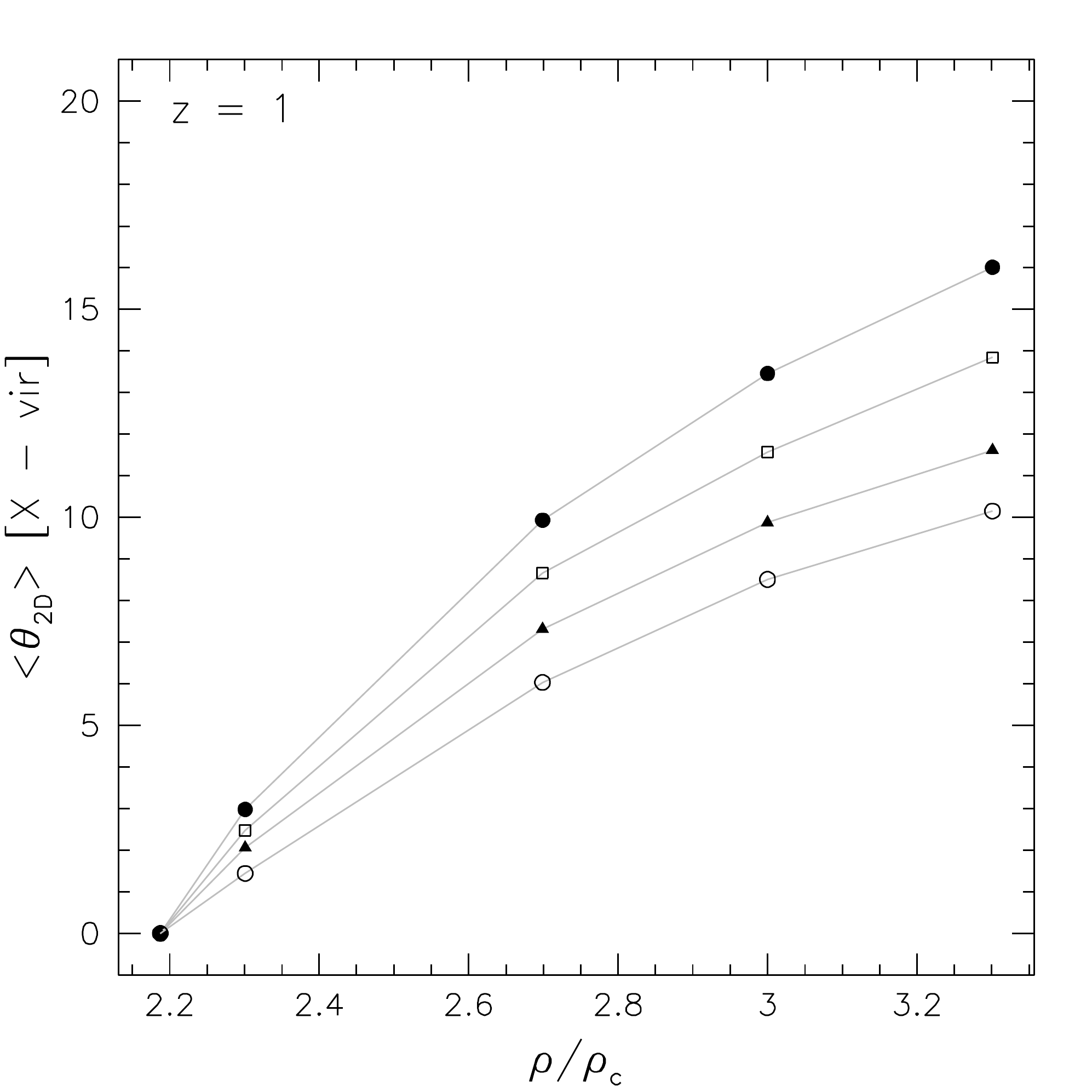}
\caption{Misalignment  of  the  ellipsoids  of  the  2D  mass  density
  distribution enclosed in different overdensities with respect to the
  direction of the  virial ellipsoid.  Left and right  panels show the
  results for  redshift $z=0$  and $z=1$, respectively.  The different
  data points indicate the misalignment  angle for different host halo
  masses as in Figure~\ref{angle1}.\label{2dalign}}
\end{figure*}

Same considerations also hold for the conditional distributions as
presented in Figure~\ref{2dcond}, where we show the 2D axis ratios at
various overdensities binning the haloes in term of the 2D axis ratios
at the virial definition, and in Figure~\ref{2dalign} which focuses on
the misalignment.

\section{Summary and Conclusions}\label{conclusions}
The aim of this  work is to provide simple and  clear estimates of how
the shape of relaxed haloes changes as a function of overdensity, mass
and redshift.  We looked at  five different ellipsoids,  enclosing various
overdensities,   and   measured  their   shape   in   three  and   two
dimensions. The main results of our work can be summarised as follows:
\begin{enumerate}
\item  general  distributions: we  confirm  that,  as found  in  other
  previous works \citep{jing02,allgood06}, dark matter haloes are more
  elongated near the centre than in  the outskirts; this is true for a
  wide    range     in    halo     masses    (from     $10^{11}$    to
  $10^{15}M_{\odot}h^{-1}$).
\item conditional  distributions: the rate  at which the  shape varies
  through the halo depends on that at the virial radius; very triaxial
  haloes show  a similar shape at  all the overdensity ellipsoids  and they are  quite well  aligned with each  other, while  for rounder
  haloes the  inner ellipsoids  are, proportionally, both  more misaligned
  and more triaxial than the virial one.
\item 2D projections: we calculated projected shapes by taking three
  different projections for each halo; the conclusions coming from
  projected quantities are similar to the 3D ones, even though the
  differences between halo ellipticities and orientations are
  shallower due to projection effects.

\end{enumerate}

Our findings  are consistent  with the  standard picture  of structure
formation,  in which  the  central  part of  haloes  may maintain  its
original triaxiality longer than the  outskirts which are subjected to
stronger interactions with the  surrounding field; also, haloes formed
recently will be  still aligned with the direction of  the last merger
or of the  filament along which matter accreted onto  the halo, and so
their whole  shape will probably  be well aligned.   The distributions
presented in this  work may be used as priors  for mass reconstruction
algorithms working  in different  wavelengths, in  order to  recover a
more realistic triaxial matter distribution of galaxies and
clusters. In this framework, it is important to keep in mind
  that haloes are not perfectly self-similar and that to reconstruct
  the virial properties of the cluster dark matter halo from the
  strong lensing a multi-parametric approach is needed.

\section*{Acknowledgments}
CG  thanks CNES  for financial  support.  ML  acknowledges the  Centre
National de  la Recherche  Scientifique (CNRS)  for its  support. This
work was performed using facilities offered by CeSAM (Centre de donneS
Astrophysique de Marseille- (http://lam.oamp.fr/cesam/). This work was
granted  access  to  the  HPC resources  of  Aix-Marseille  Universite
financed  by the  project Equip@Meso  (ANR-10-EQPX-29-01) of  the pro-
gram   \emph{Investissements  d'Avenir}   supervised  by   the  Agence
Nationale pour  la Recherche (ANR). This  work was car- ried  out with
support of  the OCEVU Labex (ANR-  11- LABX-0060) and the  A*MIDEX pro
ject  (ANR-11-IDEX-  0001-02)   funded  by  the  \emph{Investissements
  d'Avenir} French government program managed by the ANR. We also thank
  the anonymous referee for her/his useful comments that helped to improve the 
  presentation of our results.

\appendix\newpage
\section{Unrelaxed haloes}
\begin{figure} 
\includegraphics[width=\hsize]{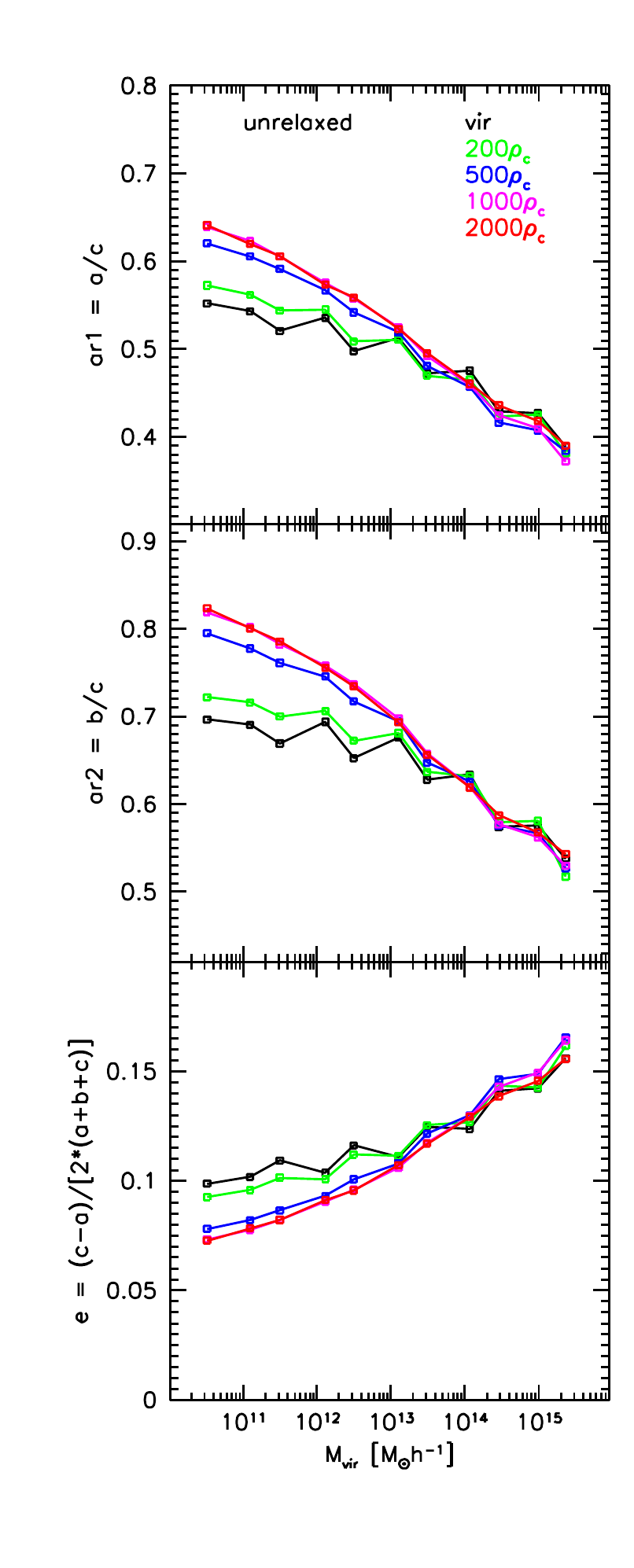}
\caption{Axial ratios and ellipticity as  a function of halo mass, for
  different overdensity thresholds for  the unrelaxed halo sample. The
  points show  the median values  of the distributions  for $ar1=a/c$,
  $ar2=b/c$    and    $e=(c-a)/[2*(a+b+c)])$   with    $a\leq    b\leq
  c$).\label{unrel}\label{lastpage}
}
\end{figure}

In the analysis preformed in our paper we have chosen to discard
unrelaxed and irregular haloes from our sample, because they tend to
introduce more scatter in the distributions, not easily to explain and
model in the same way for all the objects. In the various cases, the
analyses should take into account the specific features of each system
and of the field surrounding it.  Figure~\ref{unrel} shows the
analogous of Figure~\ref{shape_r_mass} for unrelaxed haloes, showing
clearly why they cannot be modelled together with the more relaxed
ones; in particular, the ellipticity trend is completely reversed,
leading to more triaxial shapes in the outer ellipsoids (instead of
the inner ones), probably due to the presence of multiple components
or in-falling material along a preferential direction at the observing
time.

\bibliographystyle{mn2e}
\bibliography{paper}
\end{document}